\documentclass[apj]{emulateapj}
\usepackage{lscape}
\usepackage{apjfonts}
\usepackage{graphicx}

\slugcomment{Accepted for publication in ApJ}
\shorttitle{Galactic Star Clusters in SDSS II.}
\shortauthors{An et~al.}

\begin{document}
\title{Galactic Globular and Open Clusters in the Sloan Digital Sky Survey. II.\\
Test of Theoretical Stellar Isochrones}

\author{Deokkeun An\altaffilmark{1,2},
Marc H.\ Pinsonneault\altaffilmark{1},
Thomas Masseron\altaffilmark{1},
Franck Delahaye\altaffilmark{3,4,5},\\
Jennifer A.\ Johnson\altaffilmark{1},
Donald M.\ Terndrup\altaffilmark{1,6},
Timothy C.\ Beers\altaffilmark{7},\\
Inese I.\ Ivans\altaffilmark{8,9}, and
\v{Z}eljko Ivezi\'{c}\altaffilmark{10}
}
\altaffiltext{1}{Department of Astronomy, Ohio State University,
140 West 18th Avenue, Columbus, OH 43210.}
\altaffiltext{2}{Current address: Infrared Processing and Analysis Center,
California Institute of Technology, Mail Stop 100-22, Pasadena, CA 91125;
deokkeun@ipac.caltech.edu.}
\altaffiltext{3}{Service d'Astrophysique, CEA/DSM/IRFU/SAp, CEA Saclay, 91191
Gif-sur-Yvette Cedex, France.}
\altaffiltext{4}{Centre Lasers Intenses et Applications (CELIA),
351 Cours de la Lib\'eration, 33405 Talence Cedex, France.}
\altaffiltext{5}{LERMA, Observatoire de Paris, CNRS, Universit\'e Paris Diderot,
5 place Jules Janssen, 92190 Meudon, France.}
\altaffiltext{6}{Division of Astronomical Sciences,
National Science Foundation, 4201 Wilson Blvd., Arlington, VA 22230.}
\altaffiltext{7}{Department of Physics \& Astrophysics, CSCE:
Center for the Study of Cosmic Evolution, and JINA: Joint Institute for
Nuclear Astrophysics, Michigan State University, E. Lansing, MI  48824.}
\altaffiltext{8}{The Observatories of the Carnegie Institution of
Washington, 813 Santa Barbara St., Pasadena, CA 91101.}
\altaffiltext{9}{Princeton University Observatory, Peyton Hall, Princeton, NJ 08544.}
\altaffiltext{10}{Department of Astronomy, University of Washington,
Box 351580, Seattle, WA 98195.}

\begin{abstract}
We perform an extensive test of theoretical stellar models for main-sequence
stars in $ugriz$, using cluster fiducial sequences obtained in the previous paper of this
series.  We generate a set of isochrones using the Yale Rotating Evolutionary Code (YREC)
with updated input physics, and derive magnitudes and colors in $ugriz$ from MARCS model
atmospheres.
These models match cluster main sequences over a wide range of metallicity
within the errors of the adopted cluster parameters.
However, we find a large discrepancy of model colors at
the lower main sequence ($T_{\rm eff} \la 4500$~K) for clusters at and above solar metallicity.
We also reach similar conclusions using the theoretical isochrones of Girardi et~al.\ and
Dotter et~al., but our new models are generally in better agreement with the data.
Using our theoretical isochrones, we also derive main-sequence fitting distances and turn-off
ages for five key globular clusters, and demonstrate the ability to derive these quantities from
photometric data in the Sloan Digital Sky Survey.  In particular, we exploit multiple color
indices ($g - r$, $g - i$, and $g - z$) in the parameter estimation, which allows us to
evaluate internal systematic errors.
Our distance estimates, with an error of $\sigma_{(m - M)} = 0.03$--$0.11$~mag for
individual clusters, are consistent with {\it Hipparcos}-based subdwarf
fitting distances derived in the Johnson-Cousins or Str\"omgren photometric systems.
\end{abstract}


\section{Introduction}

Improved stellar distance estimates are required to address many outstanding problems
concerning the formation and evolution of the Galaxy.
However, reliable distances from trigonometric parallaxes, such
as those from the {\it Hipparcos} mission \citep{esa:97}, are restricted to a few hundred
parsecs from the Sun.  Therefore, for now, we must rely on indirect distance measurement
techniques, such as photometric parallax or main-sequence (MS) fitting
\citep[e.g.,][]{johnson:57,eggen:59}, to probe the spatial and kinematical (sub-)structures
in the Galaxy \citep[e.g.,][]{juric:08,ivezic:08a}.

Purely empirical photometric parallax relations can be constructed 
based on trigonometric parallaxes to nearby stars.  However, the difficulty arises from
a severely restricted sample of stars with good trigonometric parallaxes over a wide range
of stellar mass and metallicity.
In particular, colors and magnitudes of MS stars are sensitive to chemical content.
Unfortunately, within the limited volume sampled by the {\it Hipparcos} mission, reliable
trigonometric parallaxes are available for only a handful of metal-poor stars
(${\rm [Fe/H] < -1.0}$), posing a challenge to Galactic halo studies relying on this information.

A widely used technique for obtaining distances to individual stars and star clusters is
to rely on theoretical predictions of stellar colors and magnitudes, tested with available
observational constraints.  Stellar evolutionary models predict
luminosities and effective temperatures ($T_{\rm eff}$) as a function of time and chemical
content (usually parameterized with [Fe/H], [$\alpha$/Fe], and helium abundance $Y$).  Stellar
atmosphere models are then used to transform theoretical quantities (luminosity and $T_{\rm eff}$)
to observables such as magnitudes and colors, expressed in terms of bolometric corrections and
color-$T_{\rm eff}$ relations.  Stellar evolutionary models can be tested against the Sun,
nearby stars with accurate parallaxes, and other stars, such as eclipsing binaries that
have accurate masses and radii.  Furthermore, multicolor photometry in nearby clusters and
field stars can be used to test bolometric corrections and color-$T_{\rm eff}$ relations
\citep[e.g.,][]{vandenberg:03,pinsono:04,an:07a,an:07b}.

\citet{pinsono:03,pinsono:04} assessed the accuracy of theoretical stellar isochrones and
reduced systematic errors in the model computation,
particularly those arising from the transformation of theoretical to observational
quantities.  They demonstrated that stellar models from the Yale Rotating Evolutionary Code
\citep[YREC;][]{sills:00} are in good agreement with observed mass-luminosity-$T_{\rm eff}$
relations for individual members ($4500 \la T_{\rm eff} {\rm (K)} \la 6000$)
in the Hyades open cluster with accurate {\it Hipparcos} parallaxes \citep{debruijne:01}.
However, they found that none of the widely-used color-$T_{\rm eff}$ relations in the
Johnson-Cousins-2MASS broadband system \citep[e.g.,][]{alonso:95,alonso:96,lejeune:97,lejeune:98} could
reproduce the observed shapes of the MS on color-magnitude diagrams (CMDs) for the Hyades open
cluster.  Since this implies problems with the adopted color-$T_{\rm eff}$ relations,
they introduced empirical corrections to the color-$T_{\rm eff}$ relations in the models.
\citet{an:07b} showed that these corrections improve the shape match to MSs of
other open clusters, as well as the internal consistency of distances from several color indices.

In this paper we perform a test of theoretical isochrones in the Sloan Digital Sky Survey
\citep[SDSS;][]{york:00,edr,dr1,dr2,dr3,dr4,dr5,dr6} $ugriz$ photometric system.
Among previous and ongoing optical surveys, SDSS is the largest and most homogeneous
database of stellar brightnesses currently available.  SDSS measures the brightnesses of
stars using a dedicated 2.5-m telescope \citep{gunn:06} in five broadband filters $u$, $g$,
$r$, $i$, and $z$, with average wavelengths of $3551$\AA, $4686$\AA, $6165$\AA, $7481$\AA, and $8931$\AA,
respectively \citep{fukugita:96,edr}.
The imaging is carried out on moonless nights of good seeing (better than
$1.6\arcsec$) under photometric conditions \citep{hogg:01}.  A portion
of the sky (along great circles) is imaged in each run by 6 columns of CCDs \citep{gunn:98}.
Astrometric positions in SDSS are accurate to
$< 0.1\arcsec$ for sources with $r < 20.5$~mag \citep{pier:03}.
However, the SDSS filters were not originally designed
for stellar observations, owing to the primarily extragalactic mission of SDSS~I.
Therefore, it is important to understand the properties of stars in this system
to exploit the full capabilities of the SDSS data set, especially for Galactic structure studies
that require accurate distance measurements.

Galactic globular and open clusters provide an ideal opportunity to achieve this goal, because
the same distance can be assumed for cluster members with a wide range of stellar masses.
However, the standard SDSS photometric pipeline \citep[{\it Photo};][]{lupton:02} failed to
provide photometry for the most crowded regions of high-density clusters.  {\it Photo} was
originally designed to handle high Galactic latitude fields with relatively low densities of
field stars, and its photometry alone does not provide sufficiently well-defined
cluster sequences.

In the previous paper of this series \citep[][hereafter Paper~I]{an:08}, we employed the
DAOPHOT/ALLFRAME \citep{stetson:87,stetson:94} suite of programs to derive
photometry for 17 globular clusters and 3 open clusters observed with SDSS.
Our DAOPHOT photometry provides well-defined cluster sequences from the lower MS to
the red giant branch (RGB).  The DAOPHOT photometry is on the native SDSS 2.5-meter system,
so cluster fiducial sequences can be directly applied to other stars observed in SDSS, without
relying upon any transformations.  This will be of particular value for the study of the space motions
of field stars in SDSS, where accurate distances are required in order to make full use of the
available proper motions.

In light of this observational improvement, we assess the accuracy of theoretical isochrones
that have been specifically constructed in the SDSS 2.5-meter photometric system.  \citet{clem:06}
compared isochrones to the fiducial sequences for three globular clusters (M13, M71, and M92)
and two open cluster (M67 and NGC~6791), in the primed ($u'g'r'i'z'$) filter system
\citep[see also][]{rider:04,fornal:07}.  However,
the $u'g'r'i'z'$ system defined by the \citet{smith:02} sample of standard stars is different
from the natural $ugriz$ system of the SDSS 2.5-m survey telescope \citep[see][]{dr1}.  Therefore, one
must rely on transformation equations \citep{tucker:06},
which limit the precision of the derived $ugriz$ magnitudes.

We use several globular and open clusters that have been the subject of extensive studies
in the literature.  Distances to most of these clusters have been estimated based on 
photometry in the Johnson-Cousins system \citep[e.g.,][hereafter C00]{reid:97,reid:98,gratton:97,carretta:00}.
Metal abundances for these clusters have also been obtained from high-resolution spectroscopy.
In particular, a recent spectroscopic abundance study of \citet[][hereafter KI03]{kraft:03}
provides an opportunity to check for possible systematic errors in cluster distance and age estimates
from the cluster abundance scale, by comparison with those from \citet[][hereafter CG97]{carretta:97}.

As a base case, we test stellar evolutionary models employed in the previous work
\citep{pinsono:03,pinsono:04,an:07a,an:07b} with updated input physics.  We further inspect
theoretical isochrones from two independent studies \citep{girardi:04,dotter:08} to check theoretical
uncertainties in the stellar interiors and atmosphere models.  Theoretical color-$T_{\rm eff}$
and bolometric corrections are also compared with those from different atmosphere models
\citep{hauschildt:99,castelli:03} to check theoretical uncertainties in the model atmospheres only.

Our goal in this study is to determine whether these models predict observed colors and magnitudes
over a wide range of metallicity.
In \S~2 we summarize the photometric accuracy of the DAOPHOT/ALLFRAME cluster photometry in Paper~I
and the cluster properties (distance, reddening, and metallicity) adopted in this study.
In \S~3 we present a new set of stellar isochrones in $ugriz$ and derive extinction coefficients.
We also compare our models with those from \citet{girardi:04} and \citet{dotter:08}.
In \S~4 we test the accuracy of these models using cluster fiducial sequences for MS stars,
and present empirical corrections on color-$T_{\rm eff}$ relations for solar and super-solar
metallicity clusters.
In \S~5 we derive MS-fitting distances and turn-off ages for our sample globular clusters,
and demonstrate the ability to derive these quantities from the SDSS photometric database,
using our set of stellar isochrones.

\section{The Cluster Data}\label{sec:data}

\subsection{The Cluster Sample}

For the calibration of photometric parallax relations in any photometric filter system,
it is important to select benchmark stellar systems with well-established stellar properties.
In particular, cluster distance, reddening, and metallicity estimates should be obtained reliably,
independently from the photometric filter system in question.  Although a few nearby open
clusters have accurate geometric distance measurements (e.g., the Hyades and the Pleiades)
in addition to precise reddening and metal abundance estimates, most of the stars in these
clusters are too bright in SDSS because of the saturation limit in the survey ($r \approx 14$~mag).

In this study, we selected six globular clusters (M3, M5, M13, M15, M71, and M92) and two open
clusters (M67 and NGC~6791) to test and calibrate theoretical stellar models.  These clusters
were included in the sample of the DAOPHOT/ALLFRAME photometric data reduction in Paper I.
For our sample open clusters, there exist reliable distances and metallicity estimates from
high-resolution spectroscopy, as discussed in more detail below.

The MS of many globular clusters in our sample are resolved $\sim2-3$~mag below the cluster
MS turn-off (MSTO).  They have spectroscopic metallicity estimates from CG97 and
KI03, which are necessary to infer the absolute magnitudes of stars more accurately.
Furthermore, distances to these clusters have been derived in the literature based on
{\it Hipparcos} parallaxes to nearby subdwarfs, either in the Johnson-Cousins or Str\"omgren
photometric systems.  We therefore view these clusters as ideal tests of the color
transformations for SDSS filters.  We first review the DAOPHOT photometry in Paper~I, and
discuss cluster properties adopted in the model comparison (\S~\ref{sec:comparison}).

\subsection{Cluster Photometry and Fiducial Sequences}

Cluster photometry was obtained in Paper I from the SDSS $ugriz$ imaging data, employing
the DAOPHOT/ALLFRAME suite of programs\footnote{Available at
{\tt http://www.sdss.org/dr6/products/\\value\_added/anjohnson08\_clusterphotometry.htm}}.
Due to the limited contamination from
background stars, fiducial sequences of our sample clusters could be accurately derived on CMDs
with several color indices ($u - g$, $g - r$, $g - i$, and $g - z$) and $r$ as a luminosity index:
hereafter, $(u - g, r)$, $(g - r, r)$, $(g - i, r)$, and $(g - z, r)$, respectively.

The zero-point error of the DAOPHOT photometry is $\sim1\%$--$2\%$, estimated
from repeated flux measurements of stars in overlapping SDSS strips/runs.  The DAOPHOT
magnitudes were tied to the {\it Photo} magnitude system using a set of cluster flanking fields,
which are far away from the dense cluster cores.  Error distributions at the bright ends indicate
photometric random errors of $\sim1\%$ in $griz$ and $\sim2\%$ in the $u$ band:
see also \citet{ivezic:04}.

In Paper~I we found that the zero points of the DAOPHOT photometry for M71 are uncertain, because of
inaccurate {\it Photo} magnitudes in the cluster's flanking fields.  For this reason, we adopted
a fiducial sequence in $u'g'r'i'z'$ from \citet{clem:08}.
We included this cluster in the following analysis because
M71 has an intermediate metal abundance (${\rm [Fe/H]} \approx -0.8$), bridging the gap between
globular and open clusters.  We converted the Clem et~al.\ fiducial sequences in the $u'g'r'i'z'$
system onto the SDSS 2.5-m $ugriz$ system, using the transformation equations in \citet{tucker:06}.

\subsection{Cluster Properties: Globular Clusters}

\begin{deluxetable}{lccccc}
\tablewidth{0pt}
\tablecaption{Sample Globular Clusters and [Fe/H] Estimates
in the Literature\label{tab:tab1}}
\tablehead{
  \colhead{Name} &
  \colhead{NGC} &
  \colhead{Harris}  &
  \colhead{\citeauthor{zinn:84}} &
  \colhead{CG97\tablenotemark{a}} &
  \colhead{KI03}
}
\startdata
M15  & 7078 & $-2.26$ & $-2.15$ & $-2.15$ & $-2.42$ \nl
M92  & 6341 & $-2.28$ & $-2.24$ & $-2.15$ & $-2.38$ \nl
M13  & 6205 & $-1.54$ & $-1.65$ & $-1.41$ & $-1.60$ \nl
M3   & 5272 & $-1.57$ & $-1.66$ & $-1.34$ & $-1.50$ \nl
M5   & 5904 & $-1.27$ & $-1.40$ & $-1.10$ & $-1.26$ \nl
M71  & 6838 & $-0.73$ & $-0.58$ & $-0.70$ & $-0.81$
\enddata
\tablenotetext{a}{Values are those shown in Table~1 of C00.}
\end{deluxetable}

Table~\ref{tab:tab1} lists [Fe/H] estimates of clusters in the literature.  In the model comparison,
we consider metallicity scales for globular clusters from CG97 and KI03 (see also Kraft \& Ivans 2004).
In each study a consistent technique was employed to derive metallicities for giants in all of our sample
clusters.  However, the CG97 values are based on \ion{Fe}{1} lines, while the KI03 [Fe/H] values
are based on \ion{Fe}{2} lines from high-resolution spectra, which were argued to be less affected by
non-local thermodynamic equilibrium (non-LTE) effects.  KI03 found that their [Fe/H] values
are about $0.2$~dex lower than those of CG97 and that they agree well with those of \citet{zinn:84} for
intermediate metallicities.

\begin{deluxetable*}{lcccccccc}
\tablewidth{0pt}
\tabletypesize{\scriptsize}
\tablecaption{Distance Moduli of Globular Clusters\label{tab:dist.gc}}
\tablehead{
  \colhead{} &
  \colhead{} &
  \multicolumn{5}{c}{{\it Hipparcos}-based Subdwarf Fitting} &
  \multicolumn{2}{c}{This Study} \\
  \cline{3-7} \cline{8-9}
  \colhead{Cluster} &
  \colhead{Harris\tablenotemark{a}} &
  \colhead{Reid (1997)} &
  \colhead{Reid (1998)} &
  \colhead{C00\tablenotemark{b}} &
  \colhead{Grundahl et~al.} &
  \colhead{KI03} &
  \colhead{${\rm [Fe/H]}_{\rm CG97}$} &
  \colhead{${\rm [Fe/H]}_{\rm KI03}$}
}
\startdata
M15 & $15.06$ & $15.38\pm0.10$ & \nodata        & \nodata         & \nodata        & $15.25$ & $15.14\pm0.11$ & $15.10\pm0.11$ \nl
M92 & $14.58$ & $14.93\pm0.10$ & \nodata        & $14.64\pm0.14$  & \nodata        & $14.75$ & $14.63\pm0.03$ & $14.66\pm0.04$ \nl
M13 & $14.42$ & $14.48\pm0.10$ & \nodata        & $14.38\pm0.13$  & \nodata        & $14.42$ & $14.41\pm0.05$ & $14.34\pm0.05$ \nl
M3  & $15.09$ & \nodata        & \nodata        & \nodata         & \nodata        & $15.02$ & $15.00\pm0.07$ & $14.95\pm0.07$ \nl
M5  & $14.37$ & $14.45\pm0.10$ & $14.52\pm0.15$ & $14.46\pm0.13$  & \nodata        & $14.42$ & $14.35\pm0.07$ & $14.26\pm0.06$ \nl
M71 & $13.02$ & \nodata        & $13.19\pm0.15$ & \nodata         & $12.86\pm0.08$ & \nodata & $12.96\pm0.08$ & $12.86\pm0.08$
\enddata
\tablenotetext{a}{Distance moduli assuming $A_V / E(B - V) = 3.1$ and reddening values from the catalog
of \citet{harris:96} February 2003 revision.}
\tablenotetext{b}{Quadrature sum of systematic and random errors.}
\end{deluxetable*}

Table~\ref{tab:dist.gc} lists distances for the globular clusters in our sample.
The second column shows distances from the catalog of \citet{harris:96} February 2003
revision, which were primarily based on the mean $V$~magnitude of the horizontal branch (HB).
The next five columns list distances derived from the {\it Hipparcos}-based subdwarf
fitting in the literature (Reid 1997, 1998; C00; Grundahl et~al. 2002; KI03).  These
studies used a subset of the {\it Hipparcos} subdwarfs to fit the MS of a cluster on CMDs,
in either the Johnson-Cousins or the Str\"omgren photometric system.  The quoted precision from
\citet{reid:97,reid:98} is shown in the table.  We computed true distance moduli of clusters
in C00, taking their adopted ratio of total to selective extinction $R_V \equiv A_V/E(B - V) = 3.1$
and $E(B - V)$.  The errors are the sum (in quadrature) of their fitting and total systematic errors
($0.12$~mag).  The sixth column shows the distance to M71 from \citet{grundahl:02},
who employed Str\"omgren $uvby$ photometry for both subdwarfs and the cluster in MS fitting.
\citeauthor{grundahl:02}
found an apparent distance modulus $(m - M)_V = 13.71\pm0.04 {\rm (ran)} \pm0.07 {\rm (sys)}$
for their adopted reddening of $E(B - V) = 0.275$, which is translated into $(m - M)_0 = 12.86\pm0.08$.
KI03 used the {\it Hipparcos} subdwarfs to derive distances to five globular clusters from
their application of the MS fitting technique.
The last two columns in Table~\ref{tab:dist.gc} list distances from MS fitting
in this study, which are discussed in \S~\ref{sec:distance}.

All of the above studies were based on a similar set of {\it Hipparcos} subdwarfs.  However,
there exist about a 10\% difference in their distance estimates.  This is because several systematic
errors were involved in the subdwarf-fitting technique, such as the metallicity scale for subdwarfs
and clusters, foreground reddening, effects of binaries, and the photometry used.  Since no study
has provided homogeneous distances to all of our sample clusters, we assumed the following
subdwarf-fitting distances in the model comparison (\S~\ref{sec:comparison}):
C00 (M5, M13, and M92), KI03 (M3 and M15), and \citet[][M71]{grundahl:02}.

We adopted the C00 distance estimates for M5, M13, and M92, since they employed the same metallicity
scale for both subdwarfs and cluster stars in the subdwarf-fitting technique.
Nevertheless, distances to M5 and M13 in C00 agree with \citet{reid:97}, \citet{reid:98}, and
KI03, within their specified errors, although these studies adopted slightly different
metallicity scales for both subdwarfs and clusters.  Notable differences can be found for M92,
where the \citet{reid:97} value is about 0.3~mag larger than the C00 estimate, and about 0.2~mag
larger than the KI03 value.  As discussed in C00, a part of this discrepancy can be explained
by the inhomogeneous metallicity scale used in \citet{reid:97} at the low metallicity end.
For this reason, we also adopted the distance to M15 from KI03.  Although \citet{grundahl:02}
utilized the Str\"omgren $uvby$ photometry to estimate the cluster metallicity, rather than
high-resolution spectroscopy, they provided an internally more consistent metallicity
scale for both subdwarfs and cluster stars than \citet{reid:98}; see Figures~10 and 11 in
\citet{grundahl:02}.

\begin{deluxetable}{lccc}
\tablewidth{0pt}
\tablecaption{Previous $E(B - V)$ Estimates for Globular Clusters\label{tab:tab11}}
\tablehead{
  \colhead{Cluster} &
  \colhead{Harris}  &
  \colhead{\citeauthor{schlegel:98}\tablenotemark{a}}  &
  \colhead{KI03}
}
\startdata
M15 & $0.10$ & $0.11$ & $0.10$ \nl
M92 & $0.02$ & $0.02$ & $0.02$ \nl
M13 & $0.02$ & $0.02$ & $0.02$ \nl
M3  & $0.01$ & $0.01$ & $0.01$ \nl
M5  & $0.03$ & $0.04$ & $0.03$ \nl
M71 & $0.25$ & $0.32$ & $0.32$
\enddata
\tablenotetext{a}{Quoted precision (16\%).}
\end{deluxetable}

Table~\ref{tab:tab11} summarizes reddening estimates for our sample globular
clusters in the literature.  The columns list values taken from the Harris
compilation, the \citet{schlegel:98} dust maps, and $E(B - V)$ estimates from
KI03, who compared colors derived from high-resolution spectroscopic determinations
of $T_{\rm eff}$ with the observed colors of the same stars.  There is no
appreciable offset in the mean between the three sets of estimates. 

Since distances from MS fitting are sensitive to reddening, the same reddening values
as in the individual subdwarf fitting studies should be used in the model comparison.
However, we adopted $E(B - V)$ estimates from KI03 for all of our sample globular clusters,
because they provided homogeneous stellar $E(B - V)$ estimates from a consistent technique.
Nevertheless, the $E(B - V)$ values adopted in C00 are only $0.005$~mag larger than
those in KI03 for M5 and M92, with the same $E(B - V)$ for M13.
In addition, with the exception of M71, the difference between the KI03 values and
those of the dust map values from \citet{schlegel:98} is $0.004\pm0.005$~mag
in the sense of \citet{schlegel:98} minus KI03.  In the case of M71,
we adopted the same $E(B - V) = 0.28$ as in \citet{grundahl:02}.  We assumed a 20\%
error in $E(B - V)$ in the model comparison (\S~\ref{sec:comparison}) for all clusters
in our sample, which encompasses the differences between the studies shown in Table~\ref{tab:tab11}.

\subsection{Cluster Properties: Open Clusters}

\begin{deluxetable*}{lccccccc}
\tablewidth{0pt}
\tabletypesize{\scriptsize}
\tablecaption{Properties of Open Clusters\label{tab:dist.oc}}
\tablehead{
  \colhead{} &
  \colhead{Alternate} &
  \colhead{High Res.} &
  \multicolumn{2}{c}{$E(B - V)$} &
  \multicolumn{2}{c}{$(m - M)_0$} &
  \colhead{} \\
  \cline{4-5} \cline{6-7}
  \colhead{NGC} &
  \colhead{Name} &
  \colhead{[Fe/H]} &
  \colhead{Previous Work} &
  \colhead{Schlegel et al.\tablenotemark{a}}  &
  \colhead{Previous Work} &
  \colhead{This Study} &
  \colhead{References}
}
\startdata
2682 & M67 & $+0.00\pm0.01$ & $0.041\pm0.004$ & $0.032$ & $ 9.61\pm0.03$ & $ 9.59\pm0.03$ & 1 \nl
6791 &     & $+0.40\pm0.03$ & $0.098\pm0.014$ & $0.155$ & $13.02\pm0.05$ & $12.95\pm0.05$ & 2
\enddata
\tablerefs{References for high resolution spectroscopic [Fe/H] values in the literature, and
estimates for stellar $E(B - V)$ and $(m - M)_0$:  (1) An et~al. 2007b, and references therein;
(2) Pinsonneault \& An (2008, in preparation), and references therein.}
\tablenotetext{a}{Quoted precision (16\%).}
\end{deluxetable*}

Table~\ref{tab:dist.oc} lists metallicity, reddening, and distance estimates for our sample
open clusters.  For M67 we took the average reddening and metallicity estimates from
high-resolution spectroscopy in the literature \citep[][and references therein]{an:07b}.
Although M67 is closer than any other cluster in this study, it is still too far away
for {\it Hipparcos} to provide a direct distance measurement.  We adopted a cluster distance
estimated from an empirically calibrated set of isochrones in the Johnson-Cousins-2MASS
system \citep{an:07b}.  In that paper, we noted that the application of these isochrones
resulted in consistent estimates for metallicity and reddening with those in the literature.

For NGC~6791 there exist four recent [Fe/H] measurements from high-resolution
spectroscopic studies:
${\rm [Fe/H]} = +0.47\pm0.04\ {\rm (random)}\pm0.08\ {\rm(systematic)}$
\citep{gratton:06}, $+0.39\pm0.01\ {\rm (random)}$ \citep{carraro:06},
$+0.35\pm0.02\ {\rm (random)}\pm0.1\ {\rm(systematic)}$ \citep{origlia:06}, and
$+0.40$ \citep{jensen:06}.
In the model comparison (\S~\ref{sec:comparison}), we adopted the average of
these measurements $\langle {\rm [Fe/H]} \rangle =+0.40\pm0.03$ with the same
weight given to each study.  Using an extended set of calibrated isochrones
in the Johnson-Cousins-2MASS system, we derived a photometric metallicity,
${\rm [Fe/H]} =+0.43\pm0.07$ (M.\ H.\ Pinsonneault \& D.\ An 2009, in preparation),
that is consistent with the above measurements, assuming solar abundance mixtures
(see Gratton et al. 2006 and Origlia et al. 2006 for subsolar [C/Fe]).
However, we found $E(B - V) = 0.098\pm0.014$ from the calibrated isochrones, while
reddening estimates in the literature vary between $0.1 \la E(B - V) \la 0.2$.
We adopted our $E(B - V)$ in the model comparison.  Both our photometric metallicity
and reddening values yield a cluster distance modulus, $(m - M)_0 = 13.02\pm0.05$,
assuming solar abundance mixtures.
If we use our adopted metallicity ${\rm [Fe/H]} =+0.40$ in this study, the distance
modulus decreases by $\approx0.03$~mag.  Given the size of the $1\sigma$ error
in distance, however, we neglected this difference, and adopted
$(m - M)_0 = 13.02\pm0.05$ for the model comparison.

\section{Models}\label{sec:model}

In this section we generate a set of theoretical isochrones in $ugriz$, which serves
as a base case for the comparison with cluster fiducial sequences (\S~\ref{sec:comparison})
and in the estimation of distances and ages of globular clusters (\S~\ref{sec:application}).
We further use other theoretical isochrones and color-$T_{\rm eff}$ relations in the
literature to inspect theoretical uncertainties in the model computations.

\subsection{Theoretical Isochrones in $ugriz$}\label{sec:ours}
 
Stellar evolutionary tracks were generated using YREC \citep{sills:00,delahaye:06} for a
wide range of compositions and ages.  Similar models were employed in our previous studies
in the Johnson-Cousins-2MASS filter system \citep{pinsono:03,pinsono:04,an:07a,an:07b},
but the new models include updated input physics (see below).  Details of
the interiors models will be presented in a paper in preparation.
 
We recently adopted atomic opacity data from the Opacity Project
\citep[OP,][]{badnell:05} for our models.  Previous models used the OPAL
opacities \citep{iglesias:96}.  In our view the OP data have benefited from 
improvements in the equation of state, especially for iron, and the availability 
of the monochromatic opacities permits tables to be readily constructed for a 
wider range of mixtures than for OPAL.  See \citet{badnell:05} for a comparison
of the underlying opacities and a discussion of the differences in the underlying
physical model.  For $T < 10^4$~K we used the molecular opacities of
\citet{alexander:94}.  We recognize that there is an inconsistency between
the mixture of heavy elements in these tables and the \citet{grevesse:98} mix adopted
in this paper, but the mixtures are quite similar, and the impact on the opacities
is typically at the 0.01-0.03 dex level for solar metallicity.  We used the OPAL~2001
equation of state \citep{rogers:02} for $T > 10^6$~K and the \citet{saumon:95}
equation of state for $T < 10^{5.5}$~K.  In the transition region between
these two temperatures, both formulations are weighted with a ramp function and
averaged. The chemical composition of each shell was updated using nuclear reaction
rates in \citet{bahcall:01}.  The models do not include gravitational settling.
A mixing-length of $\alpha = 1.78$ and the solar helium abundance $Y_\odot = 0.269$
were calibrated by matching the solar radius ($R_\odot = 6.9598\times10^8$~m) and
luminosity ($L_\odot = 3.8418\times10^{33}$~ergs s$^{-1}$) at the age of the Sun
($4.57$~Gyr); see \citet{delahaye:06}.  We used the ATLAS9 atmosphere models to
set the surface boundary condition, defined as the pressure at $\tau = 2/3$.
 
At [$\alpha$/Fe] = 0, we used the solar mixture of \citet{grevesse:98} for the initial
chemical mixture, and generated models at ${\rm [Fe/H]} = -3.0$, from ${\rm [Fe/H]} = -2.0$
to ${\rm [Fe/H]} = -0.5$ with intervals of $\Delta {\rm [Fe/H]} = 0.25$, from
${\rm [Fe/H]} = -0.5$ to ${\rm [Fe/H]} = 0.5$ with intervals of $\Delta {\rm [Fe/H]} = 0.1$,
and ${\rm [Fe/H]} = 0.75$.  We assumed the helium enrichment parameter,
$\Delta Y/ \Delta Z = 1.2$, along with the primordial helium abundance, $Y_p = 0.245 \pm 0.002$
\citep[see][]{pinsono:03}\footnote{We assumed that the helium abundance is a function of the
heavy-element content Z as given by $Y = Y_p + (\Delta Y / \Delta Z) Z$.}.  We also computed
$\alpha$-element enhanced models at [$\alpha$/Fe]$ = +0.3$ and $+0.6$ for each [Fe/H],
by increasing the ratio of $\alpha$ elements (O, Ne, Mg, Si, S, Ar, Ca, and Ti) by the
same amount in the \citet{grevesse:98} mixture using OP tables computed with the the
same mixture.  We followed the evolution of each model from pre-MS until hydrogen in the
core is exhausted.  A discussion of the post-MS evolutionary tracks is outside the scope
of this paper.
 
Theoretical isochrones in the luminosity-$T_{\rm eff}$ plane were then transformed
onto the observed CMD plane, using a set of synthetic spectra.  We used MARCS
\citep{gustafsson:08} to construct stellar atmosphere models over a wide range of
$T_{\rm eff}$ (from 4000~K to 8000~K with intervals of 250~K), $\log{g}$ (from
3.5~dex to 5.5~dex with intervals of 0.5~dex), and chemical abundances (see below).
These models are based on 1-D plane-parallel geometry, hydrostatic equilibrium, mixing
length theory (MLT) convection, and LTE.  Microturbulence ($\xi$) was chosen to be
2.0~${\rm km\ s^{-1}}$.  We assumed the same ${\rm [\alpha/Fe]}$ abundance scales
\citep{grevesse:98} as those in the interior model computations.  Based on these
models, we constructed synthetic spectra at
${\rm [Fe/H]\ ([\alpha/Fe])} =$ $-3.0\ (0.4)$, $-2.0\ (0.0,0.4)$, $-1.5\ (0.0,0.4)$,
$-1.0\ (0.0,0.4)$, $-0.5\ (0.0,0.2)$, $-0.3\ (0.0,0.1)$, $-0.2\ (0.0,0.1)$,
$-0.1\ (0.0)$, $0.0\ (0.0)$, $+0.1\ (0.0)$, $+0.2\ (0.0)$, $+0.4\ (0.0)$.

\begin{deluxetable*}{ccccccccc}
\tablewidth{0pt}
\tablecaption{Bolometric Corrections and Synthetic Colors in $ugriz$
from the MARCS Model Atmospheres\label{tab:tab}}
\tablehead{
  \colhead{[Fe/H]} &
  \colhead{[$\alpha$/Fe]} &
  \colhead{$T_{\rm eff}$} &
  \colhead{$\log{g}$} &
  \colhead{BC($r$)} &
  \colhead{$u - g$} &
  \colhead{$g - r$} &
  \colhead{$g - i$} &
  \colhead{$g - z$} \nl
  \colhead{(dex)} &
  \colhead{(dex)} &
  \colhead{(K)} &
  \colhead{(dex)} &
  \colhead{(mag)} &
  \colhead{(mag)} &
  \colhead{(mag)} &
  \colhead{(mag)} &
  \colhead{(mag)}
}
\startdata
$-3.0$ & $0.4$ & $4000$ & $3.0$ & $-0.398$ & $2.037$ & $1.115$ & $1.584$ & $1.867$ \nl
$-3.0$ & $0.4$ & $4000$ & $3.5$ & $-0.374$ & $1.988$ & $1.111$ & $1.579$ & $1.859$ \nl
$-3.0$ & $0.4$ & $4000$ & $4.0$ & $-0.354$ & $1.967$ & $1.116$ & $1.583$ & $1.860$ \nl
$-3.0$ & $0.4$ & $4000$ & $4.5$ & $-0.333$ & $1.973$ & $1.127$ & $1.592$ & $1.866$ \nl
$-3.0$ & $0.4$ & $4000$ & $5.0$ & $-0.306$ & $1.999$ & $1.146$ & $1.609$ & $1.878$ \nl
$-3.0$ & $0.4$ & $4000$ & $5.5$ & $-0.275$ & $2.057$ & $1.169$ & $1.629$ & $1.894$
\enddata
\tablecomments{The AB corrections for the SDSS magnitudes and empirical color corrections
are not applied.  Table~5 is available in its entirety in the electronic edition of the
{\it Astrophysical Journal}.  A portion is shown here for guidance regarding its form and content.}
\end{deluxetable*}

Stellar magnitudes were derived by convolving synthetic spectra with the SDSS 
$ugriz$ filter response
curves\footnote{\tt http://www.sdss.org/dr6/instruments/imager/.}, which include extinction
through an air mass of $1.3$ at Apache Point Observatory.  We integrated flux with weights given
by photon counts \citep{girardi:02}.  Magnitudes were then put onto the AB magnitude system
using a flat 3631~Jy spectrum \citep{oke:83}.  Bolometric corrections in the $r$ band and synthetic
colors are presented in Table~\ref{tab:tab}, where we adopted the bolometric magnitude of the Sun,
$M_{\rm bol,\odot} = 4.74$~mag \citep{bessell:98}.
 
In the following analysis, we adopted an $\alpha$-element enhancement scheme
motivated by the observational nature of these elements among field and cluster stars
from high-resolution spectroscopic studies, as summarized by \citet{venn:04} and 
\citet{kirby:08}:
[$\alpha$/Fe]$ = +0.4$ at [Fe/H]$ = -3.0$,
[$\alpha$/Fe]$ = +0.3$ at [Fe/H]$ = -2.0$, $-1.5$, $-1.0$,
[$\alpha$/Fe]$ = +0.2$ at [Fe/H]$ = -0.5$, and
[$\alpha$/Fe]$ = +0.0$ at [Fe/H]$ = -0.3$, $-0.2$, $-0.1$, $+0.0$, $+0.1$, 
$+0.2$, $+0.4$.
We used a linear interpolation in this metallicity grid to obtain isochrones at an
intermediate [Fe/H] value.  As an alternate case, we also tested models with [$\alpha$/Fe]$ = +0.4$
at ${\rm [Fe/H]} = -2.0, -1.5, -1.0$, given a $\Delta [\alpha{\rm/Fe}] \sim \pm0.1$ dispersion in
the observed distribution of [$\alpha$/Fe].
 
Isochrones constructed in this way are on a perfect AB magnitude system, in which magnitudes
can be translated directly into physical flux units.  However, it is known that the SDSS
photometry slightly deviates from a true AB system \citep{dr2}.  To compare our models with
fiducial sequences, we adjusted model magnitudes using AB corrections given by \citet{eisenstein:06}:
$u_{\rm AB} = u - 0.040$, $i_{\rm AB} = i + 0.015$, and $z_{\rm AB} = z + 0.030$, with no corrections
in $g$ and $r$ (see also Holberg \& Bergeron 2006).
 
The SDSS magnitudes are on the asinh magnitude system \citep{lupton:99}, which is essentially the
same as the standard astronomical magnitude at high signal-to-noise ratios.\footnote{see also
{\tt http://www.sdss.org/dr6/algorithms/fluxcal.html}.} Near the 95\% detection repeatability limit
for point sources in SDSS ($u = 22.0$~mag, $g = 22.2$~mag, $r = 22.2$~mag, $i = 21.3$~mag, and
$z = 20.5$~mag), the difference between the asinh and Pogson magnitudes are $\leq0.01$~mag, and it
rapidly becomes zero as a source becomes brighter.  Since our fiducial sequences were derived using
relatively bright stars (Paper~I), we compare theoretical isochrones on the Pogson system
directly with the DAOPHOT cluster photometry on the asinh system, without relying upon any transformation.

The isochrones constructed in this way are available at\\
{\tt http://spider.ipac.caltech.edu/~deokkeun/sdss.html}.
In the following analysis, we refer to our models as YREC+MARCS to distinguish from the other
theoretical isochrones described in \S~\ref{sec:other}.

\subsection{Extinction Coefficients}

We derived extinction coefficients in the $ugriz$ filters using MARCS synthetic spectra.
We first reddened the synthetic spectra assuming the \citet{cardelli:89} extinction
law, which is given in polynomial equations for the optical wavelength range as a function
of the wavelength and $R_V$.  We then convolved the reddened spectra with the SDSS filter
responses, over a wide range of stellar parameters and $R_V$.  In addition to the $ugriz$
filter response functions, we also used the normalized $V$ filter response from
\citet{bessell:90} to compute the absorption ratios of SDSS filters with respect to the
Johnson $V$ band, $A_\Lambda/A_V$, where $\Lambda$ represents one of the SDSS filters.

For solar metallicity models with $T_{\rm eff} = 5750$~K and $\log{g} = 4.5$, we found
$A_u/A_V = 1.567\pm0.044$,
$A_g/A_V = 1.196\pm0.013$, 
$A_r/A_V = 0.874\pm0.005$,
$A_i/A_V = 0.672\pm0.011$, and
$A_z/A_V = 0.488\pm0.014$,
assuming $R_V = 3.1$.
The errors represent the case of assuming $\Delta R_V = \pm0.2$.
Although there exists a mild variation as a function
of colors or $T_{\rm eff}$ of models, differences are within $\approx1\sigma$ of the above
estimates.  In this paper we neglected these color-dependent variations in extinction ratios,
since most of the sample clusters have interstellar reddening less than $E(B - V) \sim 0.1$.

\citet{girardi:04} also estimated extinction coefficients based on the same reddening law
\citep{cardelli:89}, but using synthetic ATLAS9 spectra.  Our values are in good agreement
with their values.  At $4000 \leq T_{\rm eff} {\rm (K)} \leq 8000$ with $\log{g} = 4.5$,
the differences are only $\Delta A_{\Lambda}/A_V \sim 0.01$.

We additionally derived extinction coefficients in $u'g'r'i'z'$ for the fiducial sequence
of M71 \citep{clem:08}, since transformations from
$ugriz$ to $u'g'r'i'z'$ were derived for unreddened colors \citep{tucker:06}.
We derived extinction coefficients for USNO $u'g'r'i'z'$ filters\footnote{
See {\tt http://www-star.fnal.gov/ugriz/Filters/\\response.html}.}
using solar metallicity models with $T_{\rm eff} = 5750$~K and $\log{g} = 4.5$ at $R_V = 3.1$:
$A_{u'}/A_V = 1.569$,
$A_{g'}/A_V = 1.177$,
$A_{r'}/A_V = 0.866$,
$A_{i'}/A_V = 0.656$,
and $A_{z'}/A_V = 0.488$.

For a given $R_V$ and $E(B - V)$ of a cluster, we computed extinction in the $r$ band and
color-excess values in $ugriz$ from
\begin{equation}
A_r = \left( \frac{A_r}{A_V} \right) E(B - V) R_V,
\end{equation}
and
\begin{equation}
E(\Lambda_1 - \Lambda_2) =
\left( \frac{A_{\Lambda_1}}{A_V} - \frac{A_{\Lambda_2}}{A_V} \right) E(B - V) R_V,
\end{equation}
where $\Lambda_1$ and $\Lambda_2$ represent each of the SDSS filter passbands.

\subsection{Other Models: Isochrones}\label{sec:other}

Isochrones in $ugriz$ are also available from two other independent studies.
\citet{girardi:04} provided the first extensive set of isochrones in $ugriz$ filters,
based on evolutionary models used in their earlier studies \citep[see][]{girardi:02}.
To derive bolometric corrections and color-$T_{\rm eff}$ relations, they have mainly
used ATLAS9 non-overshooting models \citep{castelli:97,bessell:98}; hereafter we
refer to these isochrones as Padova+ATLAS9.  On the other hand, \citet{dotter:08}
recently presented the Dartmouth Stellar Evolution Program (DSEP) isochrones, where
PHOENIX model atmospheres \citep{hauschildt:99} were employed to convert luminosity
and $T_{\rm eff}$ into magnitudes and colors; hereafter we refer to these
isochrones as DSEP+PHOENIX.  In this section we compare different evolutionary models
($L-T_{\rm eff}$) and atmosphere models (bolometric corrections and color-$T_{\rm eff}$
relations) separately.

\begin{figure*}
\epsscale{1.0}
\plotone{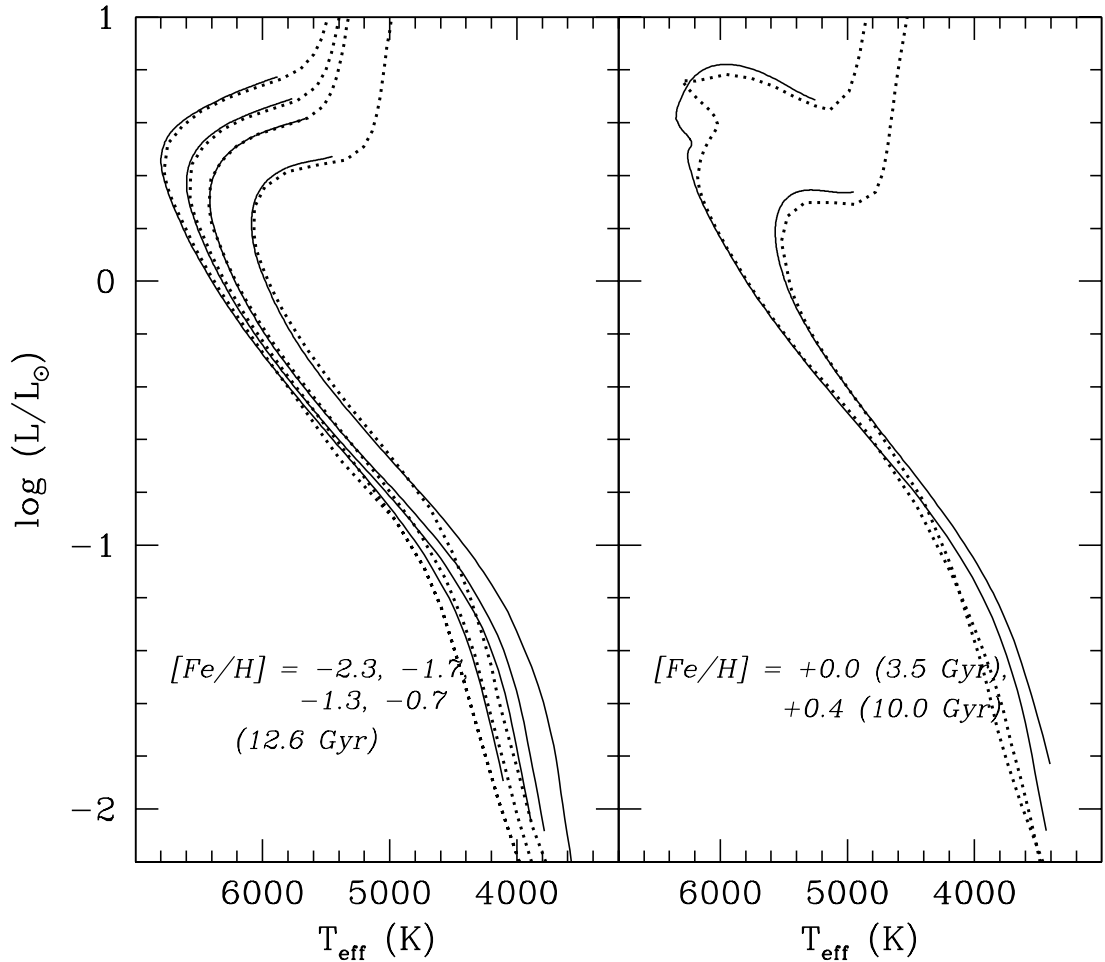}
\caption{Comparison between Padova ({\it dotted line}) and YREC ({\it solid line})
evolutionary models.  {\it Left:} Isochrones are shown at [Fe/H] = $-2.32$, $-1.71$, $-1.31$,
and $-0.71$, with [$\alpha$/Fe] = $0.0$.  Ages of $12.6$~Gyr are assumed.
{\it Right:} Isochrones are shown at [Fe/H] = $0.0$ ($3.5$~Gyr) and $+0.37$ ($10.0$~Gyr) with
[$\alpha$/Fe] = $0.0$.\label{fig:comp.2}}
\end{figure*}

Figure~\ref{fig:comp.2} shows the comparison between Padova ({\it dotted lines}) and
YREC ({\it solid}) isochrones on the $\log(L/L_\odot)-T_{\rm eff}$ plane.  Padova isochrones
are shown at $Z = 0.0001$ (${\rm [Fe/H]} = -2.32$), $Z = 0.0004$ (${\rm [Fe/H]} = -1.71$),
$Z = 0.001$ (${\rm [Fe/H]} = -1.31$), $Z = 0.004$ (${\rm [Fe/H]} = -0.71$),
$Z = 0.019$ (${\rm [Fe/H]} = +0.00$), and $Z = 0.040$ (${\rm [Fe/H]} = +0.37$), taking
their adopted solar composition ($Y = 0.273$ and $Z = 0.019$) to compute [Fe/H] values.
All of these models are based on
the solar abundance ratio, since \citet{girardi:04} do not provide models with $\alpha$-element
enhancement.

As shown in Figure~\ref{fig:comp.2}, YREC and Padova isochrones are generally in good
agreement with each other, but they exhibit a large difference at the lower MS,
in the sense that the Padova models are hotter than the YREC models for a given luminosity.
This difference is likely due to the use of different equations of state at low temperatures.
Padova models use those in \citet{mihalas:90}, while YREC models adopt the
\citet{saumon:95} equation of state.  The morphologies of the MSTOs differ
for the solar metallicity, 3.5~Gyr models, because Padova models include convective
core overshooting \citep[see][]{girardi:00}.

\begin{figure*}
\epsscale{1.0}
\plotone{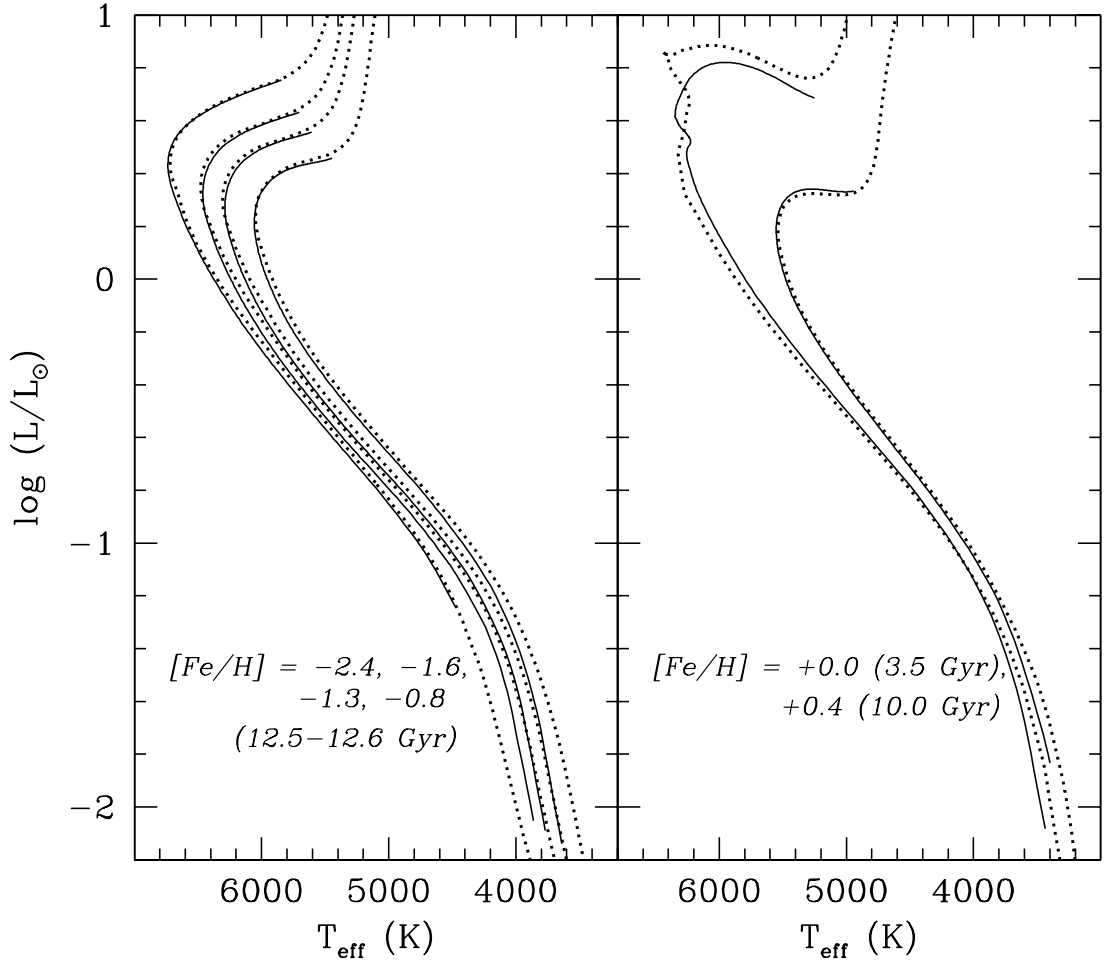}
\caption{Comparison between DSEP ({\it dotted line}) and YREC ({\it solid line})
evolutionary models.  {\it Left:} Isochrones are shown at [Fe/H] = $-2.42$, $-1.60$, $-1.26$,
and $-0.81$, with $\alpha$-element enhancement adopted in this study
([$\alpha$/Fe]$ \approx +0.3$ at [Fe/H]$ \la -1$; see \S~\ref{sec:ours}).
Ages of $12.6$~Gyr are assumed.
{\it Right:} Isochrones are shown at [Fe/H] = $0.00$ ($3.5$~Gyr) and $+0.40$ ($10.0$~Gyr)
with [$\alpha$/Fe] = $0.0$.\label{fig:comp}}
\end{figure*}

Figure~\ref{fig:comp} shows a similar comparison in the $\log(L/L_\odot)-T_{\rm eff}$ plane
between DSEP ({\it dotted line}) and YREC ({\it solid line}), both of which include
$\alpha$-element enhancement.  DSEP isochrones at intermediate [$\alpha$/Fe] values were
obtained by interpolating models over their customized grid at [$\alpha$/Fe]$ = 0.0$, $0.2$,
and $0.4$.  In the left-hand panel, these isochrones are shown at
${\rm [Fe/H]} = -2.42$, $-1.60$, $-1.26$, and $-0.81$, with our adopted $\alpha$-element
enhancement scheme ([$\alpha$/Fe]$ \approx +0.3$ at [Fe/H]$ \la -1$; see \S~\ref{sec:ours}).
In the right-hand panel, isochrones are shown at [Fe/H]$ = +0.0$ and $+0.4$ with no
$\alpha$-element enhancement.

As shown in Figure~\ref{fig:comp}, the two models are generally in good agreement with each other
on the $\log(L/L_\odot)$ vs.\ $T_{\rm eff}$ plane.  In fact, DSEP has been developed based on
a similar code to YREC \citep{demarque:07}, and many input ingredients are the same as in
YREC.  However, they have been subsequently modified by the two independent groups.  For example,
DSEP includes the PHOENIX model atmosphere to set the surface boundary condition, while YREC
adopts the ATLAS9 model.  DSEP uses high temperature opacities from OPAL \citep{iglesias:96},
while YREC utilizes OP data \citep{badnell:05}.  As mentioned above, the different MSTO
morphology of 4~Gyr, solar-metallicity isochrones is due to the inclusion/exclusion of
the convective core overshoot in DSEP/YREC.

\subsection{Other Models: Bolometric Corrections and Color-$T_{\rm eff}$ Relations}\label{sec:other2}

\begin{figure}
\epsscale{1.2}
\plotone{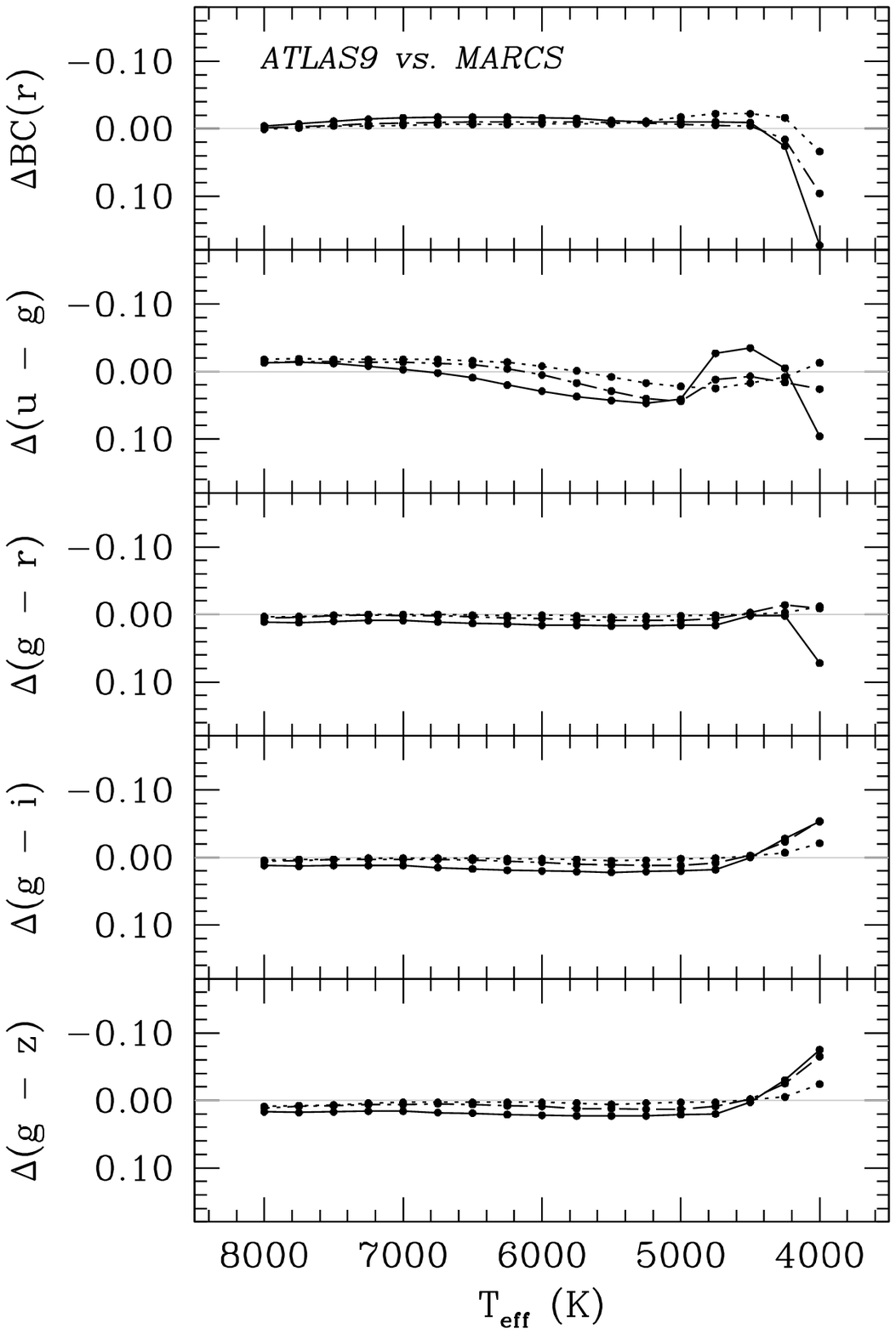}
\caption{Comparisons in $r$-band bolometric corrections and synthetic colors between ATLAS9 and
MARCS atmosphere models, in the sense of ATLAS9 minus MARCS.  Comparisons are shown at
$\log{g} = 4.0$, with three different metallicities:
[Fe/H]$ = -2.0$ and [$\alpha$/Fe]$ = 0.4$ ({\it dotted line}),
[Fe/H]$ = -1.0$ and [$\alpha$/Fe]$ = 0.4$ ({\it dot-dashed line}),
[Fe/H]$ = +0.0$ and [$\alpha$/Fe]$ = 0.0$ ({\it solid line}).
\label{fig:comp.atlas}}
\end{figure}

Figure~\ref{fig:comp.atlas} compares $r$-band bolometric corrections and synthetic
colors from ATLAS9\footnote{{\tt http://wwwuser.oat.ts.astro.it/castelli/grids.html}.} and MARCS,
in the sense of the former minus the latter.  We derived magnitudes and colors for ATLAS9 spectra,
following the same steps as for MARCS.  Comparisons are shown at
$\log{g} = 4.0$ with three different metallicities:
[Fe/H]$ = -2.0$ and [$\alpha$/Fe]$ = 0.4$ ({\it dotted line}),
[Fe/H]$ = -1.0$ and [$\alpha$/Fe]$ = 0.4$ ({\it dot-dashed line}),
[Fe/H]$ = +0.0$ and [$\alpha$/Fe]$ = 0.0$ ({\it solid line}).
Although the difference between ATLAS9 and MARCS is becoming larger at a lower $T_{\rm eff}$ end
in the comparison, the agreement between these two models is generally good ($\la0.02$~mag in
bolometric corrections and colors) at $5000 \la T_{\rm eff} {\rm (K)} \la 8000$.  The comparison
in $u - g$ suggests a large theoretical uncertainty in colors that involve the $u$ bandpass.

\begin{figure}
\epsscale{1.2}
\plotone{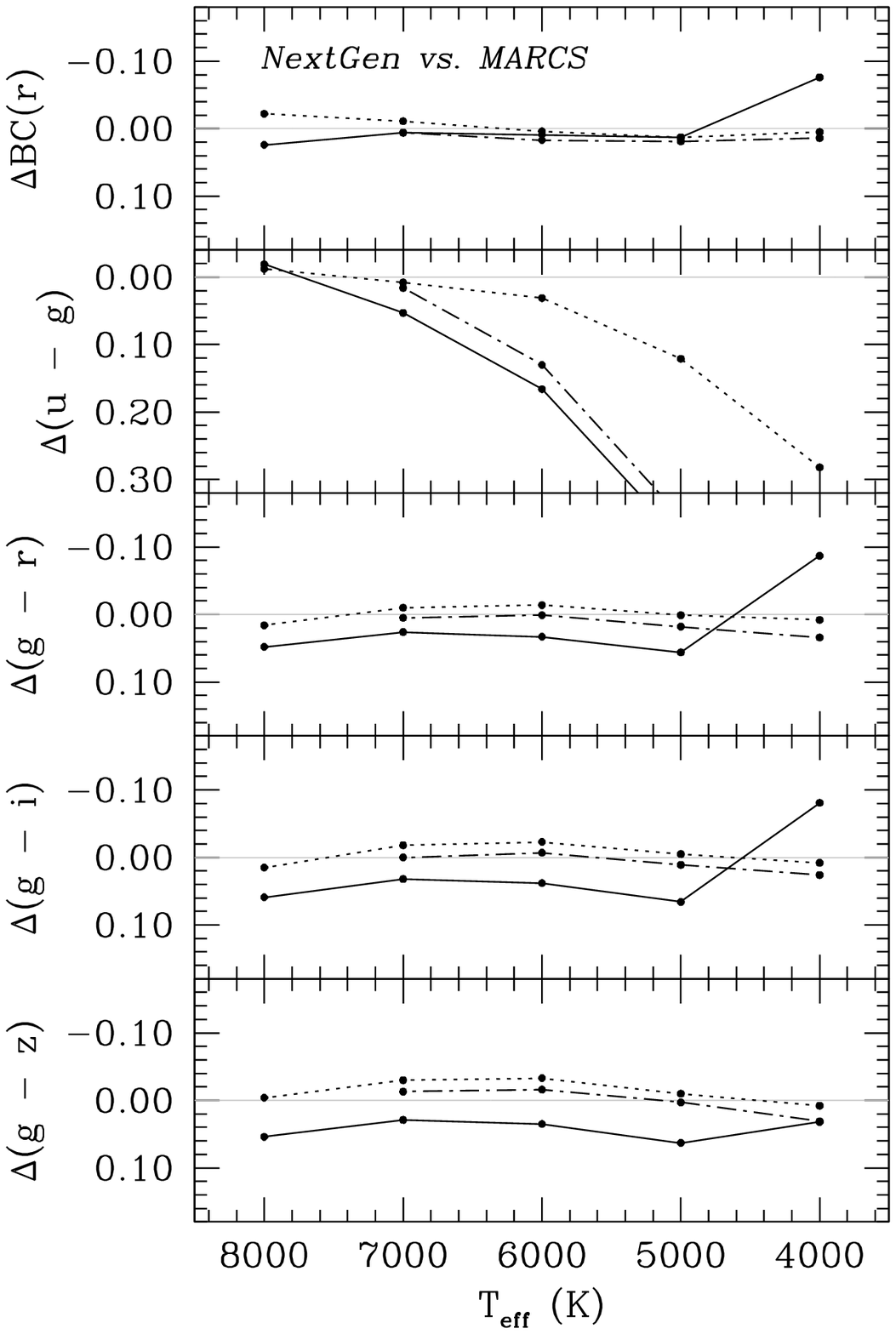}
\caption{Comparisons in $r$-band bolometric corrections and synthetic colors between
PHOENIX/NextGen and MARCS atmosphere models, in the sense of PHOENIX/NextGen minus MARCS.
Comparisons are shown at $\log{g} = 4.0$, with three different metallicities: [Fe/H] $ = -2.0$
({\it dotted line}), $-1.0$ ({\it dot-dashed line}), and $+0.0$ ({\it solid line}).
All models assume the solar abundance ratio.
\label{fig:comp.nextgen}}
\end{figure}

Similarly, Figure~\ref{fig:comp.nextgen} shows comparisons in $r$-band bolometric corrections and
synthetic colors between MARCS and the LTE-based NextGen grid generated with the PHOENIX
code.\footnote{{\tt http://phoenix.ens-lyon.fr/Grids/NextGen/SPECTRA/}.}
Comparisons are shown at $\log{g} = 4.0$ with three different metallicities:
[Fe/H] $ = -2.0$ ({\it dotted line}), $-1.0$ ({\it dot-dashed line}), and $+0.0$ ({\it solid line}).
However, $\alpha$-element enhanced models were not included in this comparison,
because they were not available in the public PHOENIX database.  The differences are in the sense
of PHOENIX/NextGen minus MARCS.  Unlike ATLAS9 model colors, PHOENIX/NextGen models have
larger color differences from MARCS in all color indices, especially in $u - g$.

\section{Model Comparison}\label{sec:comparison}

In this section, we assess the accuracy of theoretical isochrones using cluster
fiducial sequences for our sample clusters.  For our base case, we use YREC+MARCS
models to check whether stellar models satisfactorily reproduce observed colors
and magnitudes of stars over a wide range of metal abundance.  In this comparison,
we take into account various systematic errors, such as those from the photometric
calibration, cluster distance, and the metallicity scale for globular clusters.
We also test Padova+ATLAS9 and DSEP+PHOENIX isochrones with cluster fiducial sequences.

\subsection{Systematic Errors in Color-$T_{\rm eff}$ Relations}

\begin{figure*}
\epsscale{1.0}
\plotone{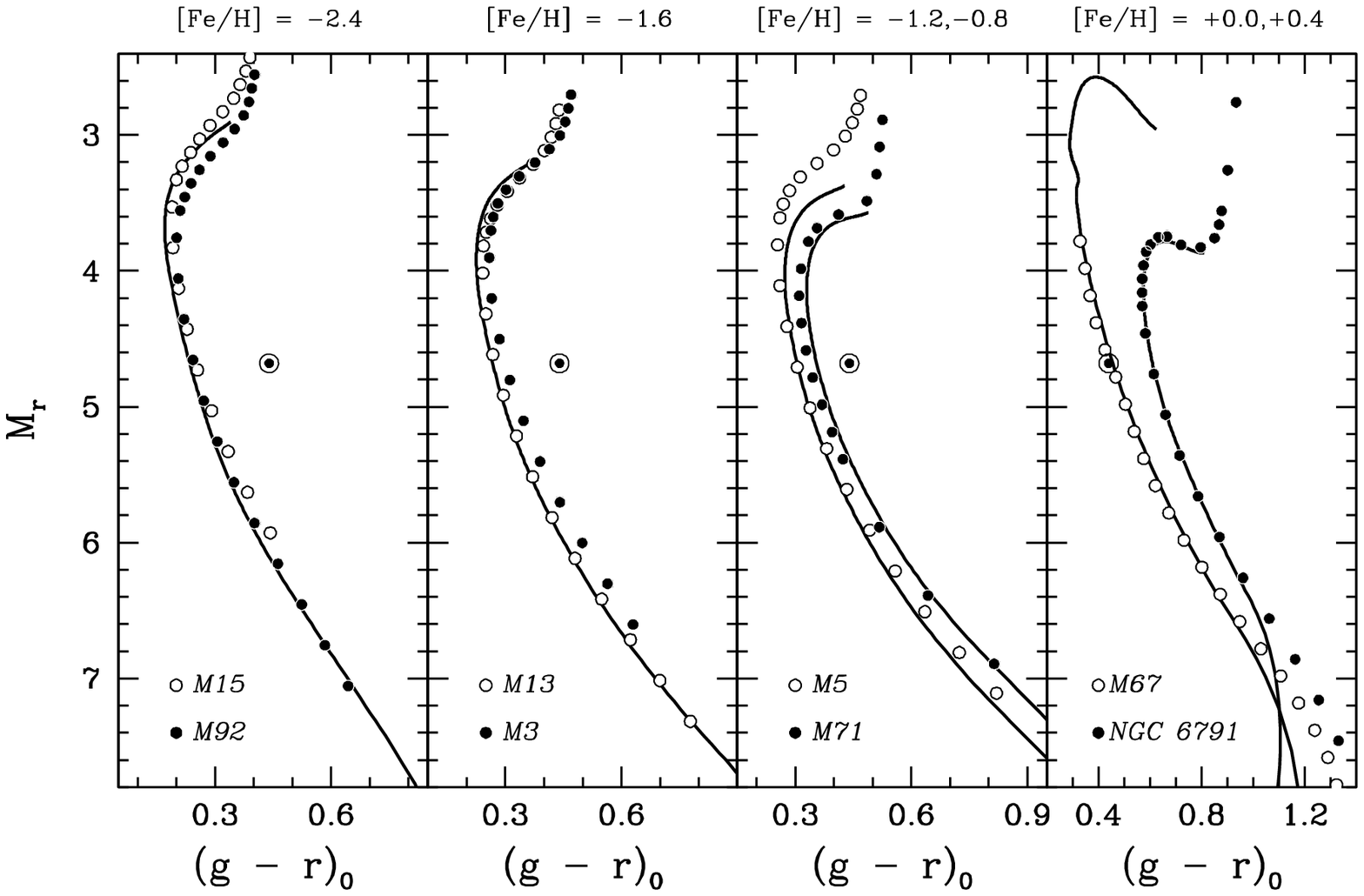}
\caption{Reddening corrected $(g - r, M_r)$ CMDs for sample clusters.  Open and closed circles are
fiducial points for each cluster.  {\it Left three panels}: Solid lines are YREC+MARCS
models at the age of 12.6~Gyr with ${\rm [Fe/H]} = -2.4, -1.6, -1.2, -0.8$, as indicated on
top of each panel.  Distances from the $Hipparcos$-based subdwarf fitting are assumed, with
canonical cluster reddening values (see text).  The fiducial points for M71 are
from Clem et~al., after transforming $u'g'r'i'z'$ into $ugriz$ using transformation
equations in Tucker et~al.  {\it Right panel}: Solid lines are models with
[Fe/H] = $+0.0$ (4.0~Gyr), $+0.3$ (10.0~Gyr).  Bull's-eyes mark the approximate position of
the Sun, to guide eyes on the metallicity sensitivity of colors and magnitudes.
\label{fig:all.gr}}
\end{figure*}

\begin{figure*}
\epsscale{1.0}
\plotone{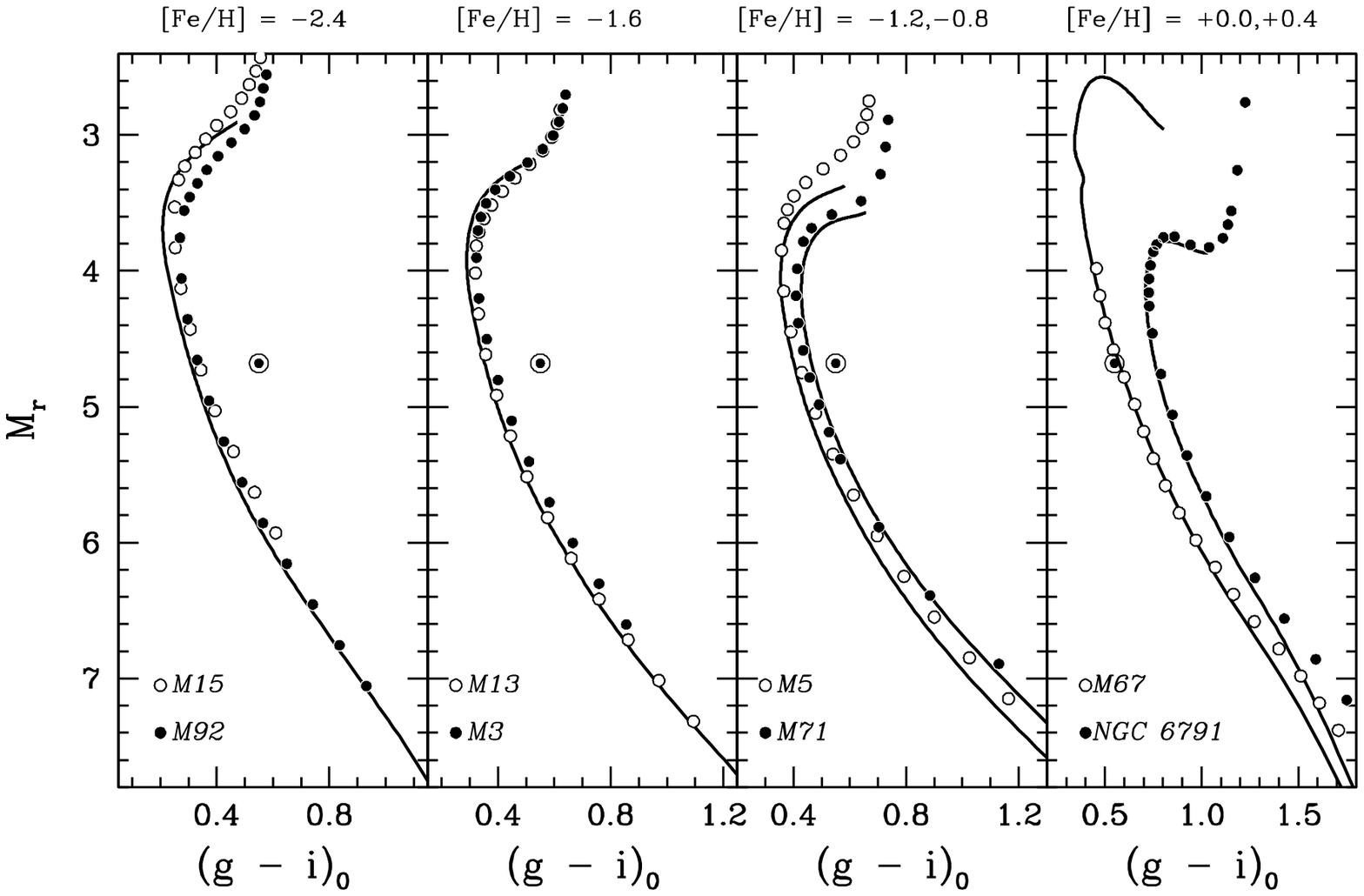}
\caption{Same as in Fig.~\ref{fig:all.gr}, but $(g - i, M_r)$.
\label{fig:all.gi}}
\end{figure*}

\begin{figure*}
\epsscale{1.0}
\plotone{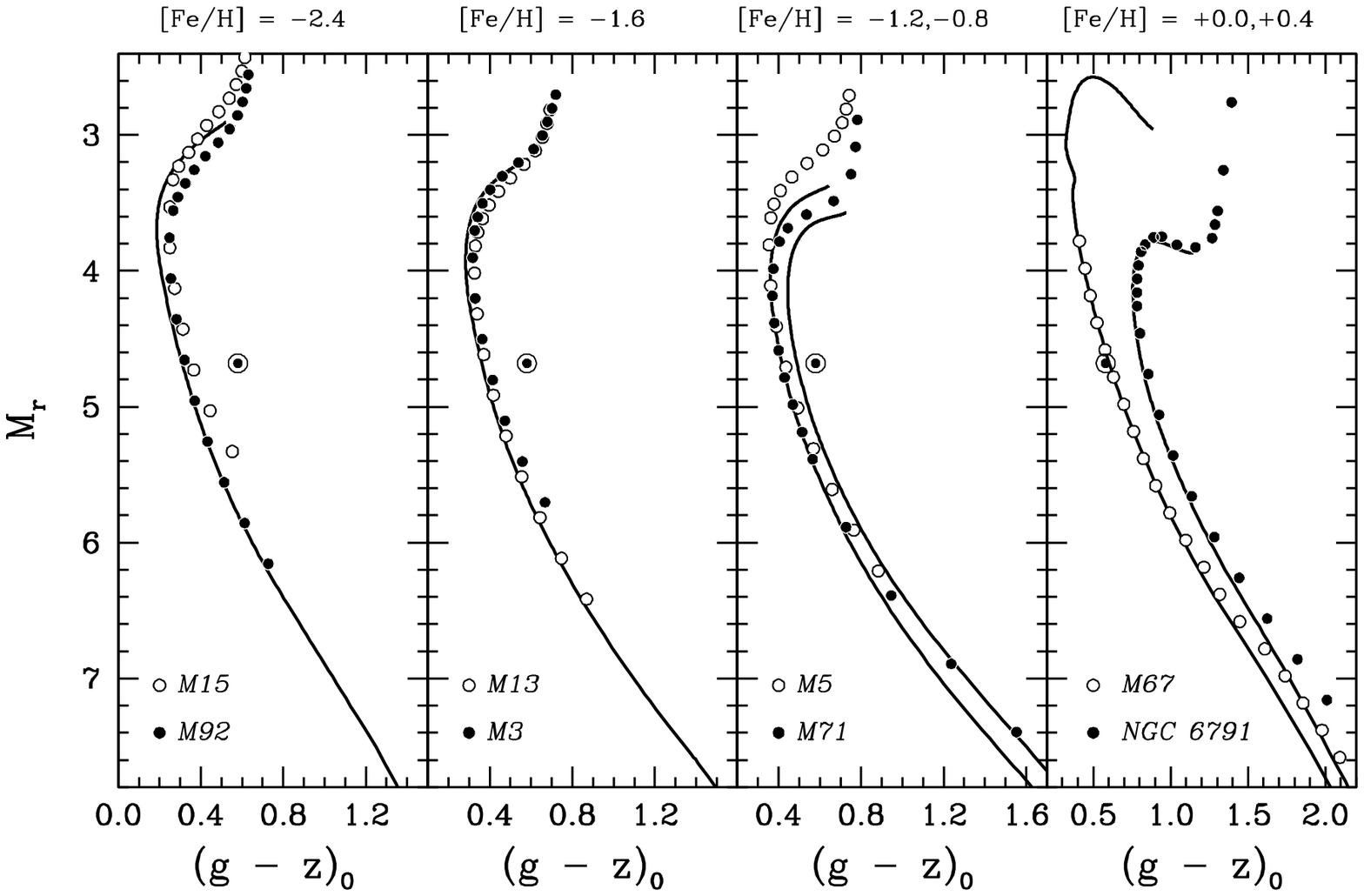}
\caption{Same as in Fig.~\ref{fig:all.gr}, but $(g - z, M_r)$.
\label{fig:all.gz}}
\end{figure*}

\begin{figure*}
\epsscale{1.0}
\plotone{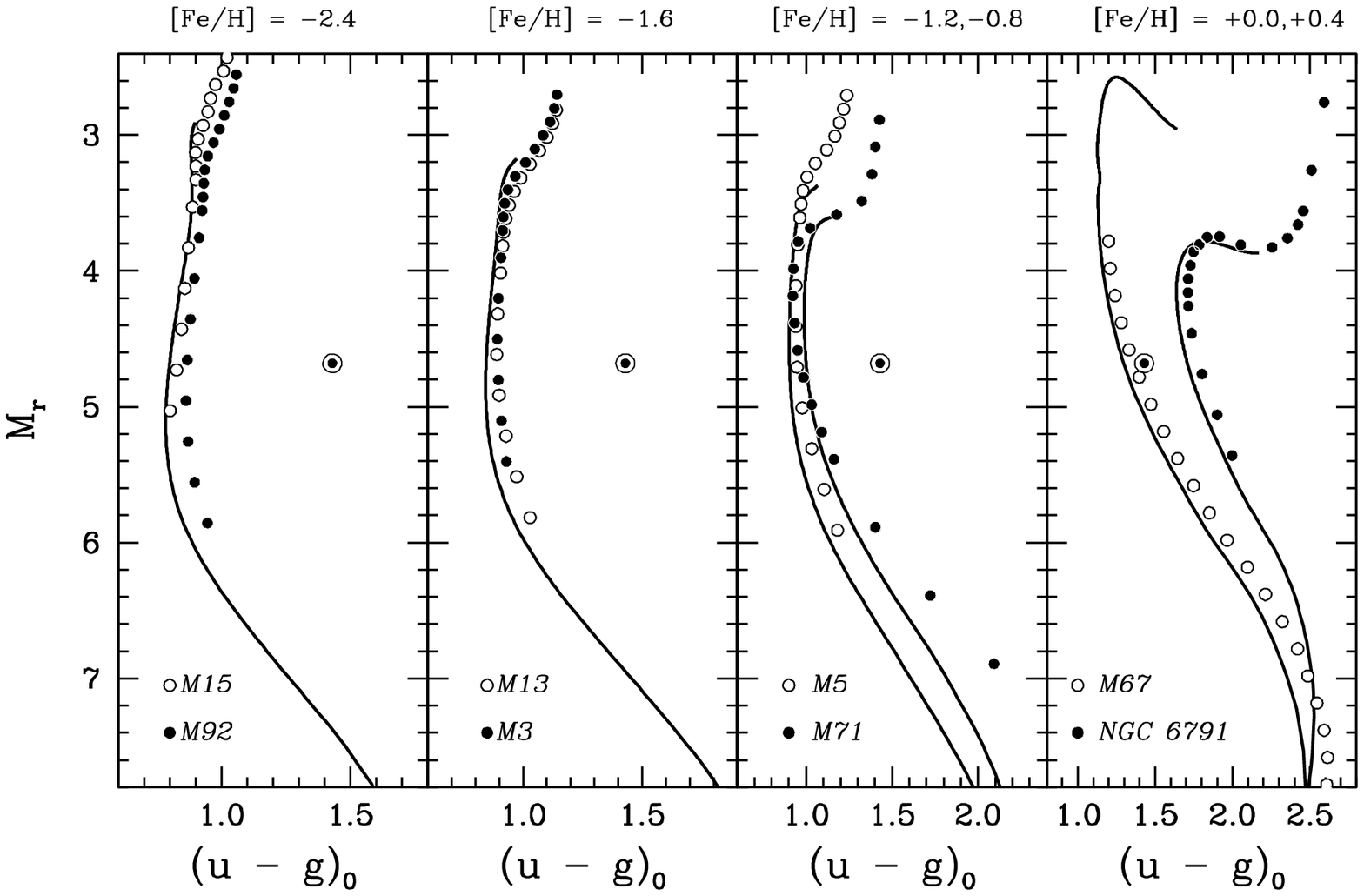}
\caption{Same as in Fig.~\ref{fig:all.gr}, but $(u - g, M_r)$.
\label{fig:all.ug}}
\end{figure*}

Figures~\ref{fig:all.gr}--\ref{fig:all.ug} show comparisons between YREC+MARCS models
and fiducial sequences in $(g - r, r)$, $(g - i, r)$, $(g - z, r)$, and $(u -g, r)$,
respectively.  In each panel, fiducial sequences are displayed as either filled or open
circles, and the models are shown as solid lines with [Fe/H] as indicated on top of
each panel.  The leftmost panel contains fiducial sequences for the two most
metal-poor clusters (M15 and M92).  The second and third panels display those for
intermediate-metallicity globular clusters (M3, M5, M13, and M71).  The last panel
displays fiducial sequences for two open clusters (M67 and NGC~6791).  To place
fiducial sequences on dereddened color vs.\ absolute magnitude diagrams, we adopted
the cluster distance and reddening values as discussed in \S~\ref{sec:data}.  The
range of $M_r$ covered by a fiducial sequence is different from one cluster to the
other, because SDSS images were taken in drift-scan or time-delay-and-integrate (TDI)
mode, with a fixed effective exposure time of $54.1$ seconds per source.  For the models,
we assumed ages of 12.6~Gyr for all globular clusters, 3.5~Gyr for M67, and 10.0~Gyr
for NGC~6791.  To guide the eye, the approximate positions of the
Sun\footnote{See {\tt http://www.sdss.org/dr6/algorithms/\\sdssUBVRITransform.html}.}
are shown on the CMDs as bulls-eyes.

\begin{figure}
\epsscale{1.2}
\plotone{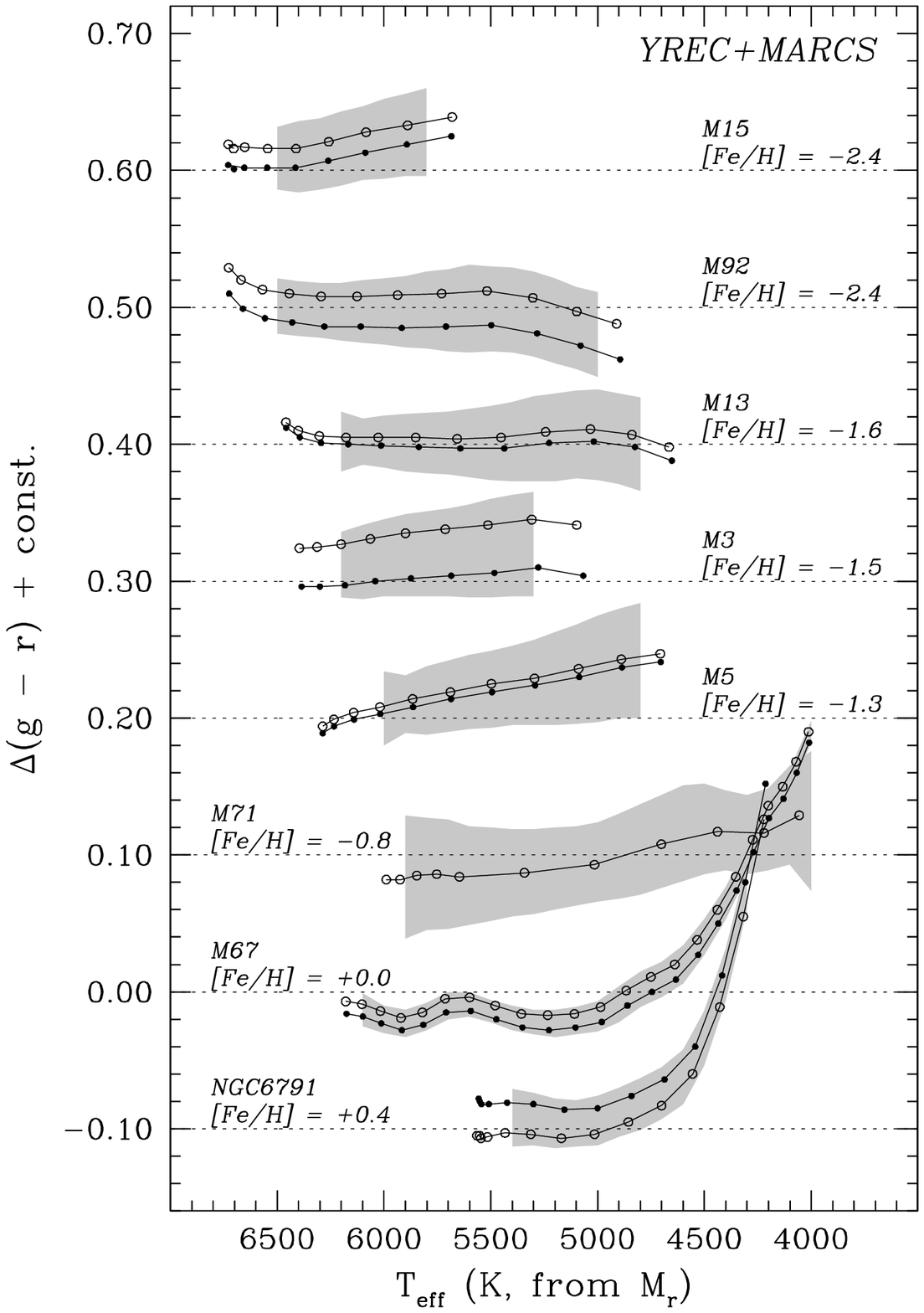}
\caption{Color differences in $g - r$ between YREC+MARCS models and fiducial sequences.
The KI03 [Fe/H] values are assumed with the {\it Hipparcos}-based subdwarf fitting distances
for globular clusters.  Individual cluster cases are shown in such a way that the most
metal-poor cluster is shown at the top, while the most metal-rich open cluster is shown at the bottom,
with an offset $\Delta (g - r) = 0.10$ between them.  Dotted lines indicate zero differences
between model and the data.  For each cluster, filled and open circles
exhibit the size of a systematic error from the photometric calibration (see text).  A grey strip
represents a $\pm1\sigma$ range of a total systematic error in the comparison.  The result for M71
is based on Clem et~al.\ fiducial sequence (after a photometric transformation to $ugriz$).
\label{fig:cteff.gr}}
\end{figure}

\begin{figure}
\epsscale{1.2}
\plotone{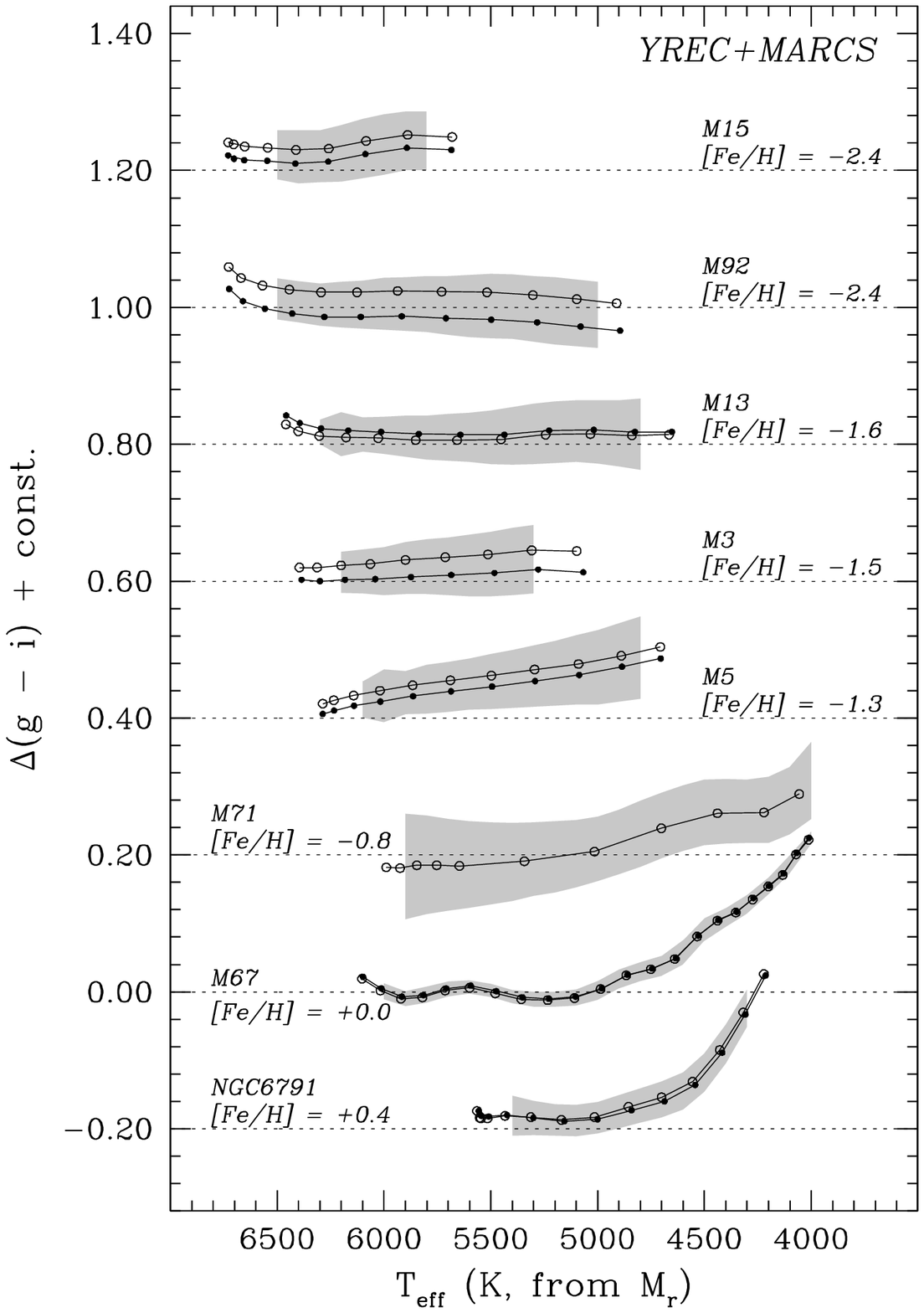}
\caption{Same as in Fig.~\ref{fig:cteff.gr}, but model comparisons in $g - i$.
\label{fig:cteff.gi}}
\end{figure}

\begin{figure}
\epsscale{1.2}
\plotone{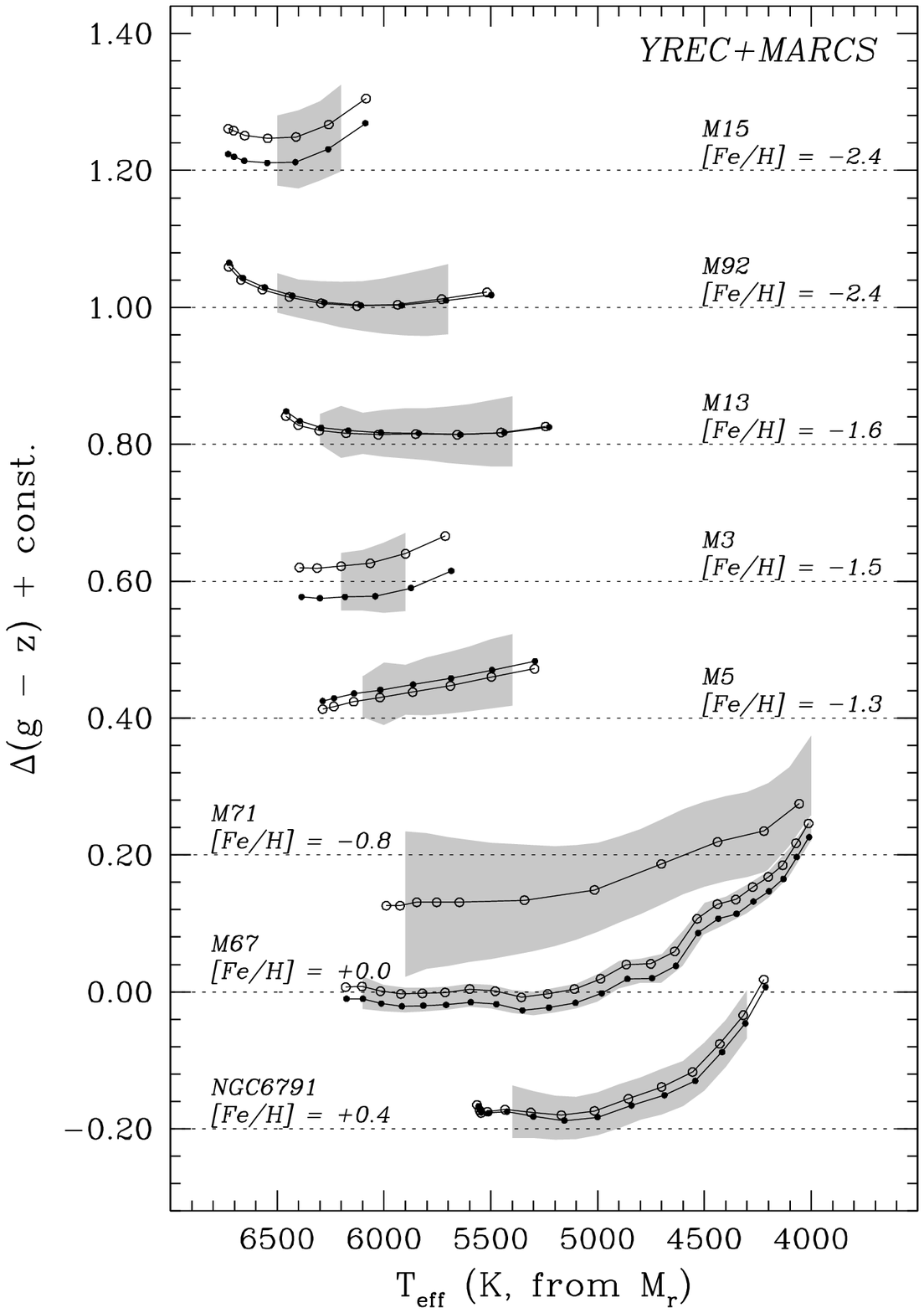}
\caption{Same as in Fig.~\ref{fig:cteff.gr}, but model comparisons in $g - z$.
\label{fig:cteff.gz}}
\end{figure}

\begin{figure}
\epsscale{1.2}
\plotone{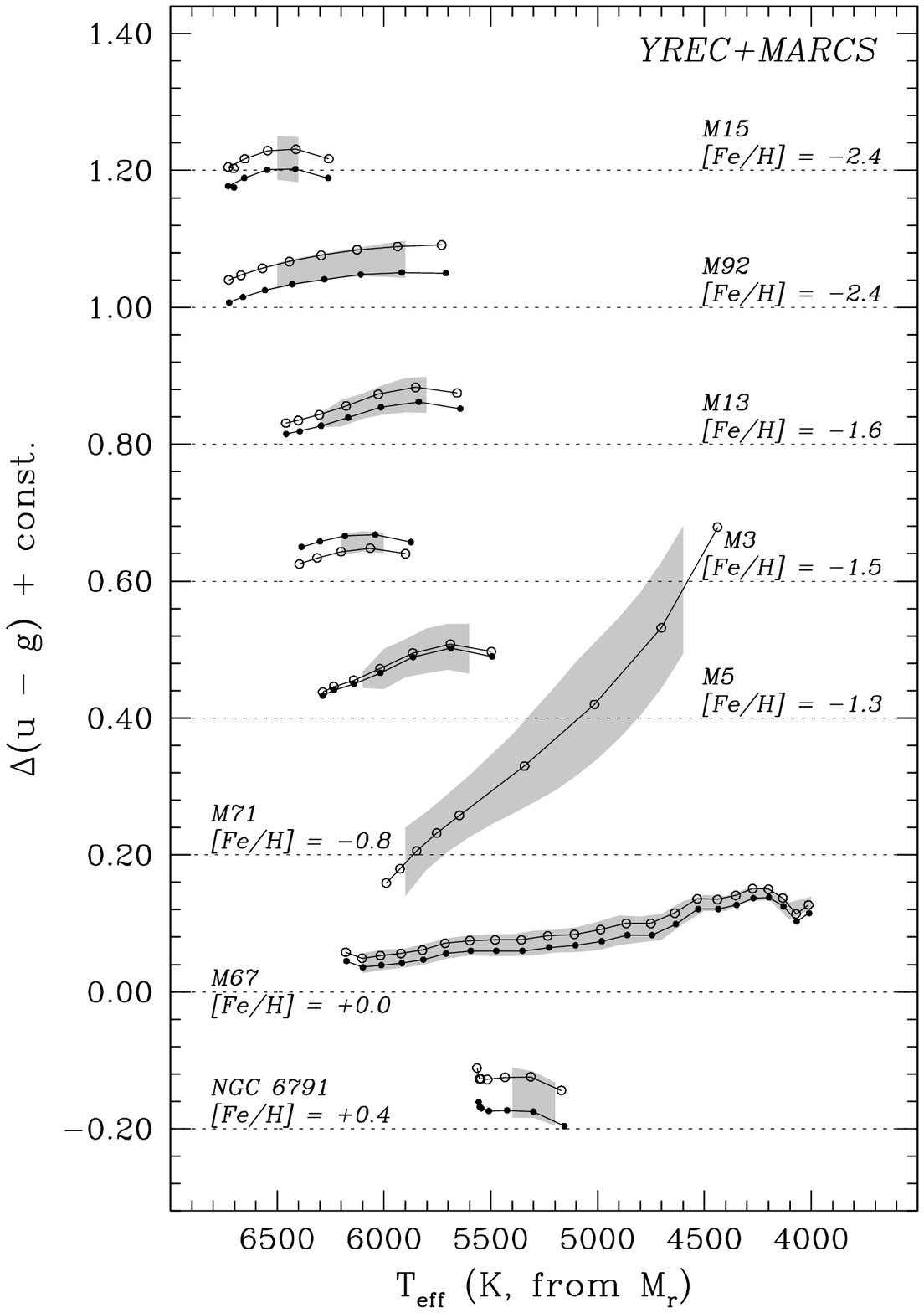}
\caption{Same as in Fig.~\ref{fig:cteff.gr}, but model comparisons in $u - g$.
\label{fig:cteff.ug}}
\end{figure}

The model comparisons in Figures~\ref{fig:all.gr}--\ref{fig:all.ug} are reproduced
in Figures~\ref{fig:cteff.gr}--\ref{fig:cteff.ug} as a function of $T_{\rm eff}$ in
$g - r$, $g - i$, $g - z$, and $u - g$, respectively.  We estimated color differences
between models and cluster fiducial sequences at a given $M_r$, which was then converted
into $T_{\rm eff}$ in the model.  Open circles represent color differences for each of
the fiducial points, as shown in Figures~\ref{fig:all.gr}--\ref{fig:all.ug}.  In this
comparison we used models at the KI03 [Fe/H] values for the globular clusters.
Dotted lines indicate zero differences between model and the data for each cluster.

In Figures~\ref{fig:cteff.gr}--\ref{fig:cteff.ug} the difference between open and filled
circles effectively show the size of a systematic error in the model comparison from
photometric zero-point errors.  In Paper~I we used {\it Photo} magnitudes in a set of
flanking fields, which are far from the crowded cluster field, to set the zero point of
DAOPHOT cluster photometry.  This procedure was necessary to put DAOPHOT cluster
photometry on the {\it Photo} magnitude scale.  There were typically two SDSS imaging
runs covering each cluster, and DAOPHOT photometry was independently calibrated on each
run.  However, about $1\%$--$2\%$ systematic differences were found in DAOPHOT photometry
for overlapping stars in these two runs.  For this reason, we combined photometry from
two runs by selecting a ``reference'' run, which was simply chosen based on the sky
coverage of a cluster or the run number.  Open circles in
Figures~\ref{fig:cteff.gr}--\ref{fig:cteff.ug} and the data in
Figures~\ref{fig:all.gr}--\ref{fig:all.ug} show the case when we adopt these zero points
for our sample clusters.  On the other hand, filled circles show cases when using an
alternate set of zero points for the same stars.  In most instances, this is the zero
point established by the other run.  For M67 and NGC~6791, there were images taken in three
SDSS runs, but we only display the case that shows the largest difference in magnitudes.
In M71, open circles show the model comparison with the \citet{clem:08} fiducial sequence
(after a photometric transformation to $ugriz$), where we adopted a 0.01~mag zero-point
error in each passband (see below).

The grey strip in Figures~\ref{fig:cteff.gr}--\ref{fig:cteff.ug} represents a $\pm1\sigma$
range in the color comparison from various systematic errors.  We assumed the difference
between open and filled circles as a $2\sigma$ error from the photometric calibration.
We adopted quoted errors in distance from individual studies (\S~\ref{sec:data}).  Since
KI03 did not provide the distance error, we assumed $\sigma_{(m - M)} = 0.15$ mag, which
is a reasonable size for the errors of the {\it Hipparcos}-based subdwarf fitting (e.g.,
Reid 1997, 1998; C00).  Note that KI03 derived cluster distances from eyeball fitting on
$(B - V, V)$, did not correct for a bias due to binaries (for both the cluster and
subdwarf samples), and also did not apply the Lutz-Kelker corrections in their fitting
procedure.  We assumed $\Delta R_V = \pm0.2$ for the error in the reddening laws, and took
a $20\%$ error in $E(B - V)$ and in the cluster age ($\Delta \log{t}~{\rm (Gyr)} \approx 0.1$).
We estimated the error from [$\alpha$/Fe] by examining the case of [$\alpha$/Fe]$ = 0.4$ at
${\rm [Fe/H]} < 1.0$ (\S~\ref{sec:model}).

\begin{deluxetable*}{lccccc}
\tablewidth{0pt}
\tablecaption{Systematic Errors in the Comparison of Model Colors\label{tab:sys2}}
\tablehead{
  \colhead{Source of Error} &
  \colhead{$\Delta$ Quantity\tablenotemark{a}} &
  \colhead{$\Delta (g - r)$} &
  \colhead{$\Delta (g - i)$} &
  \colhead{$\Delta (g - z)$} &
  \colhead{$\Delta (u - g)$}
}
\startdata
Fitting Residual        & \nodata    & $\pm0.003$ & $\pm0.004$ & $\pm0.006$ & $\pm0.008$\nl
Phot.\ Calibration      & $1\%-2\%$  & $\pm0.007$ & $\pm0.007$ & $\pm0.008$ & $\pm0.011$\nl
$(m - M)_0$             & $\pm0.15$  & $\pm0.020$ & $\pm0.029$ & $\pm0.031$ & $\pm0.017$\nl
$R_V$                   & $\pm0.2$   & $\pm0.002$ & $\pm0.002$ & $\pm0.001$ & $\pm0.006$\nl
$E(B - V)$              & $20\%$     & $\pm0.006$ & $\pm0.012$ & $\pm0.019$ & $\pm0.008$\nl
$\log{t}$               & $20\%$     & $\pm0.003$ & $\pm0.005$ & $\pm0.009$ & $\pm0.008$\nl
$[\alpha/{\rm Fe}]$\tablenotemark{b} & $\pm0.1$ & $\pm0.002$ & $\pm0.003$ & $\pm0.003$ & $\pm0.002$\nl
${\rm [Fe/H]}$\tablenotemark{b}      & $\pm0.1$ & $\pm0.007$ & $\pm0.009$ & $\pm0.009$ & $\pm0.014$\nl
Total                   & \nodata    & $\pm0.024$ & $\pm0.034$ & $\pm0.040$ & $\pm0.029$
\enddata
\tablecomments{Values are shown for the average color difference over
$4500 \leq T_{\rm eff} {\rm (K)} \leq 6500$.}
\tablenotetext{a}{Approximate size of the errors.  See \S~\ref{sec:data} for details.}
\tablenotetext{b}{Open cluster data are not included in the error estimation.}
\end{deluxetable*}

Systematic errors in the comparison of model colors are listed in Table~\ref{tab:sys2}.
They are the average values estimated from all of our sample clusters.  To avoid a large
discrepancy in the models at the lower MS for M67 and NGC~6791 (see below), we took
the average in the color difference over $4500 \leq T_{\rm eff} {\rm (K)} \leq 6500$ for
each cluster.  In addition to the individual error sources described above, we also included
an error from the fitting residual.  Even after shifting isochrones in colors, we often
found that they do not perfectly match the shape of the fiducial sequence.  We computed
this as an error in the average color difference between the model and the cluster fiducial
sequence.  We discuss the effect of the globular cluster [Fe/H] scale in more detail
in the next section.  The total error is the quadratic sum of all of the error
contributions, which is dominated by the error in the cluster distance.

As shown in Figures~\ref{fig:cteff.gr}--\ref{fig:cteff.gz}, there is no compelling evidence that
model colors differ from the observed MS for globular clusters in $g - r$, $g - i$, and $g - z$.
On the other hand, there is a strong departure of the models at the lower MS for two
open clusters ($T_{\rm eff} \la 4500$~K).  The models predict bluer colors than the observed ones
in all color indices.  In fact, the problem has long been noted in other photometric systems,
such as the Johnson-Cousins system \citep[e.g.,][]{sandquist:04}, for many contemporary
low-mass stellar models.  We present empirical corrections for the models in \S~\ref{sec:emp}.

There is an $\sim0.1$~mag offset in $u - g$ for all clusters as shown in
Figure~\ref{fig:cteff.ug}, with no apparent trend in metallicity.  Since the other color indices
based on the $g$ bandpass exhibit good agreement with the data, the $u - g$ color offset suggests
a problem in the $u$ bandpass.  The $\sim5\%$ color difference is larger than expected from
the photometric calibration error in SDSS.  Instead, the difference is most likely due to
missing opacities in model atmospheres at short wavelengths.  The difference is in the sense
that models overestimate the flux in the $u$ bandpass.  The comparison in $u - g$ for M71 is
worse than those for the other clusters in our sample, suggesting a problem in the color
transformation for this highly reddened cluster.

\subsection{Metallicity Scale for Globular Clusters}\label{sec:fehscale}

\begin{figure}
\epsscale{1.2}
\plotone{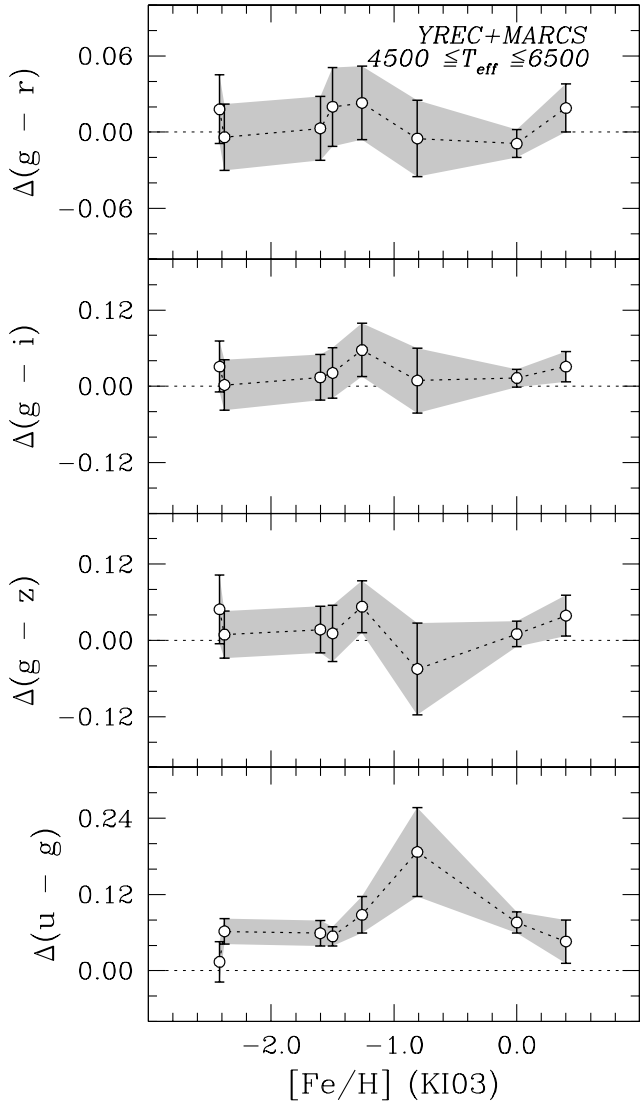}
\caption{Average color difference between the models and fiducial sequences in $ugriz$.
Open circles represent individual clusters, showing an average color difference
computed at $4500 \leq T_{\rm eff} {\rm (K)} \leq 6500$.  For globular clusters the
KI03 [Fe/H] values are used.  Error bars and grey strips represent $\pm1\sigma$ range
of a total systematic error.
\label{fig:average.fiducial}}
\end{figure}

Figure~\ref{fig:average.fiducial} shows the color differences between YREC+MARCS
models and fiducial sequences, as a function of cluster metallicity, on the KI03 scale.
Each open circle represents an average color difference for individual clusters over
$4500 \leq T_{\rm eff} {\rm (K)} \leq 6500$.  The grey strip displays a $\pm1\sigma$
error in Table~\ref{tab:sys2}, excluding the error from [Fe/H].  As seen in the figure,
model colors are generally bluer than fiducial sequences for globular clusters.  The
average color differences are
$\Delta \langle g - r \rangle = +0.008\pm0.011$,
$\Delta \langle g - i \rangle = +0.022\pm0.017$,
$\Delta \langle g - z \rangle = +0.021\pm0.018$, and
$\Delta \langle u - g \rangle = +0.059\pm0.009$ for
the five globular clusters in our sample.

\begin{figure}
\epsscale{1.2}
\plotone{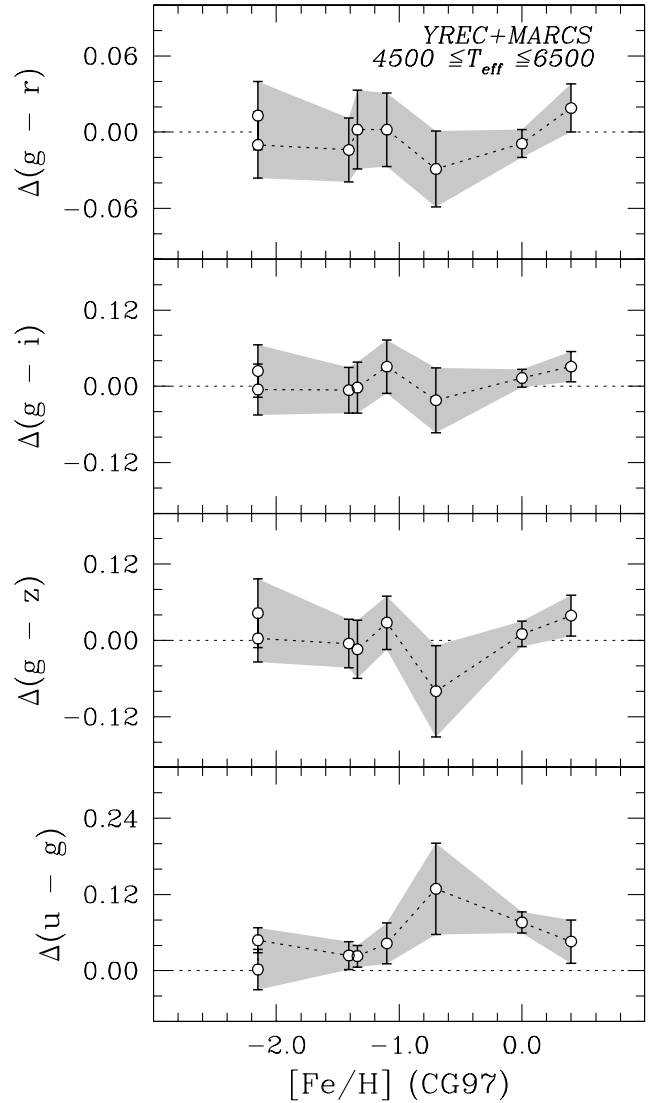}
\caption{Same as in Fig.~\ref{fig:average.fiducial}, but when the CG97 [Fe/H] values
are used in the model comparison for globular clusters.
\label{fig:average.feh}}
\end{figure}

Figure~\ref{fig:average.feh} shows a model comparison when the CG97 [Fe/H] values are used.
In comparison to the KI03 case (Fig.~\ref{fig:average.fiducial}), color differences
between our models and fiducial sequences are reduced by $\sim0.02$~mag:
$\Delta \langle g - r \rangle = -0.006\pm0.011$,
$\Delta \langle g - i \rangle = +0.004\pm0.017$,
$\Delta \langle g - z \rangle = +0.002\pm0.018$, and
$\Delta \langle u - g \rangle = +0.031\pm0.010$.
As discussed in KI03, their [Fe/H] values are about $0.2$~dex lower than those from CG97,
which make the isochrones redder.

We took the difference from the two [Fe/H] scales as an effective $2\sigma$ error, which
is listed in Table~\ref{tab:sys2}.  However, the metallicity scale has a relatively small
impact on the model comparison, since the total error budget is still dominated by the
error in the subdwarf-fitting distances.  In the above analyses, we did not take a random error
in [Fe/H] ($\sigma_{\rm \langle [Fe/H] \rangle} \approx \pm0.01$--$0.02$) into account, 
as it is much smaller than the systematic error ($\sigma_{\rm \langle [Fe/H] \rangle}
\approx \pm0.1$).  We note that both CG97 and KI03 were based on spectroscopic measurements
of red giants, which could be different from those for dwarfs.  However, the difference may
not be larger than $\sim0.1$~dex \citep[e.g.,][]{korn:06,korn:07}.

\subsection{Tests of Other Models}

\begin{figure}
\epsscale{1.2}
\plotone{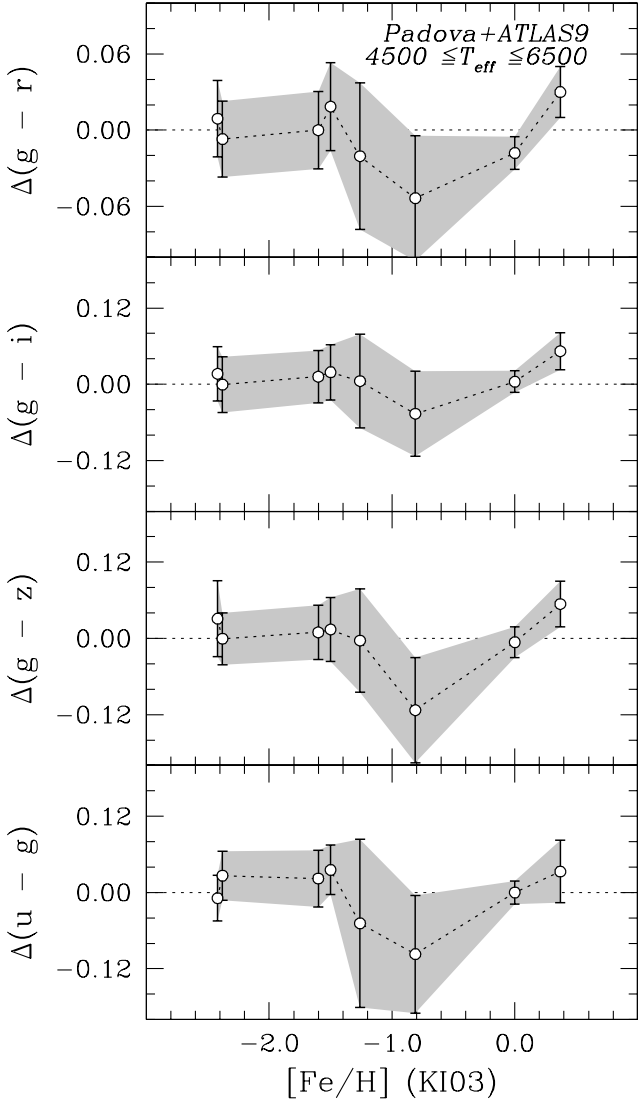}
\caption{Same as in Fig.~\ref{fig:average.fiducial}, but for Padova+ATLAS9 models.  For each
globular cluster, isochrones with two different [Fe/H] values are used in the model comparison
to bracket the observed cluster abundance (see text), because Padova+ATLAS9 models do not include
$\alpha$-element enhancement.
\label{fig:average.padv}}
\end{figure}

Figure~\ref{fig:average.padv} shows a color comparison between cluster fiducial sequences and
Padova+ATLAS9 isochrones.  Because \citet{girardi:04} did not include $\alpha$-element enhanced
models, we used their models in the comparison at the following [Fe/H] values, in order to bracket the
observed [Fe/H] abundance of a cluster \citep[e.g.,][]{salaris:93,kim:02}:
${\rm [Fe/H]} = -2.3$ and $-1.7$ for M15 and M92,
${\rm [Fe/H]} = -1.7$ and $-1.3$ for M3 and M13,
${\rm [Fe/H]} = -1.3$ and $-0.7$ for M5,
and ${\rm [Fe/H]} = -0.7$ and $-0.4$ for M71.
For the open clusters, we used their models at ${\rm [Fe/H]} = +0.0$ for M67, and
${\rm [Fe/H]} = +0.37$ for NGC~6791.  We applied AB corrections for SDSS magnitudes
(\S~\ref{sec:model}).
As shown in Figure~\ref{fig:average.padv}, Padova+ATLAS9 models describe the observed
cluster data reasonably well.  However, the error bars (grey strip) are larger than those for
the YREC+MARCS cases (Figs.~\ref{fig:average.fiducial} and \ref{fig:average.feh}), because
of the wider range of [Fe/H] values adopted in the model comparison.

\begin{figure}
\epsscale{1.2}
\plotone{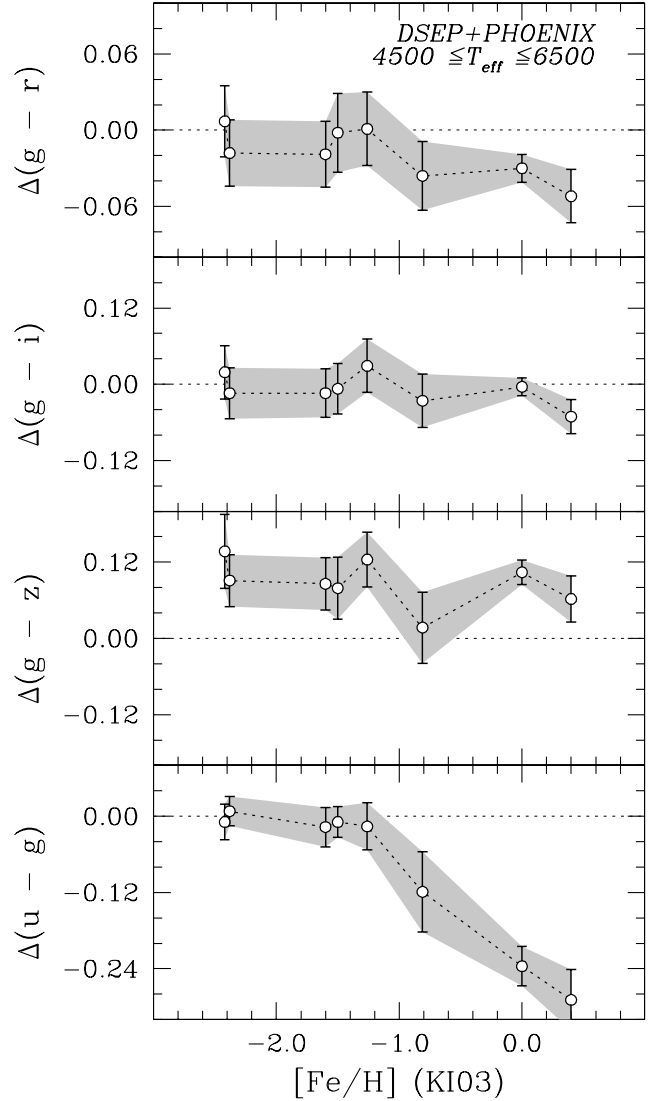}
\caption{Same as in Fig.~\ref{fig:average.fiducial}, but for DSEP+PHOENIX models.
The KI03 [Fe/H] values are used for globular clusters.
\label{fig:average.dsep.fiducial}}
\end{figure}

Similarly, Figure~\ref{fig:average.dsep.fiducial} shows a comparison between DSEP+PHOENIX
models and fiducial sequences.  Since their models are available over a wide range of [Fe/H]
and [$\alpha$/Fe], isochrones were interpolated at the cluster's [Fe/H] on the KI03 scale.
Although model colors and magnitudes from DSEP+PHOENIX already included AB corrections,
$0.01$~mag was subtracted from their $g$-band magnitudes, to be consistent with our adopted
AB corrections for SDSS.  Although DSEP+PHOENIX models are in reasonably good agreement
with the cluster data, they show a large difference in $g - z$ and $u - g$ from the data.
The above tests suggest that YREC+MARCS models are generally in better agreement with
observed cluster data than these models.

\subsection{Empirical Corrections on Color-$T_{\rm eff}$ Relations at [Fe/H]$\ga0$}\label{sec:emp}

In the above two sections, we restricted our model comparison to $T_{\rm eff} \geq 4500$~K
because the models are in poor agreement with the lower MS at solar and super-solar
metallicity (Fig.~\ref{fig:cteff.gr}--\ref{fig:cteff.ug}).  Both the Padova+ATLAS9 and
DSEP+PHOENIX isochrones show similar discrepancies in colors for open clusters.  Since CMDs
for globular clusters do not extend far below $T_{\rm eff} \sim 5000$~K, similar problems
could not be found at the lower MS.  However, the comparison with M71 suggests that
model colors and magnitudes are probably accurate at ${\rm [Fe/H]} \la -0.8$.

\begin{figure}
\epsscale{1.2}
\plotone{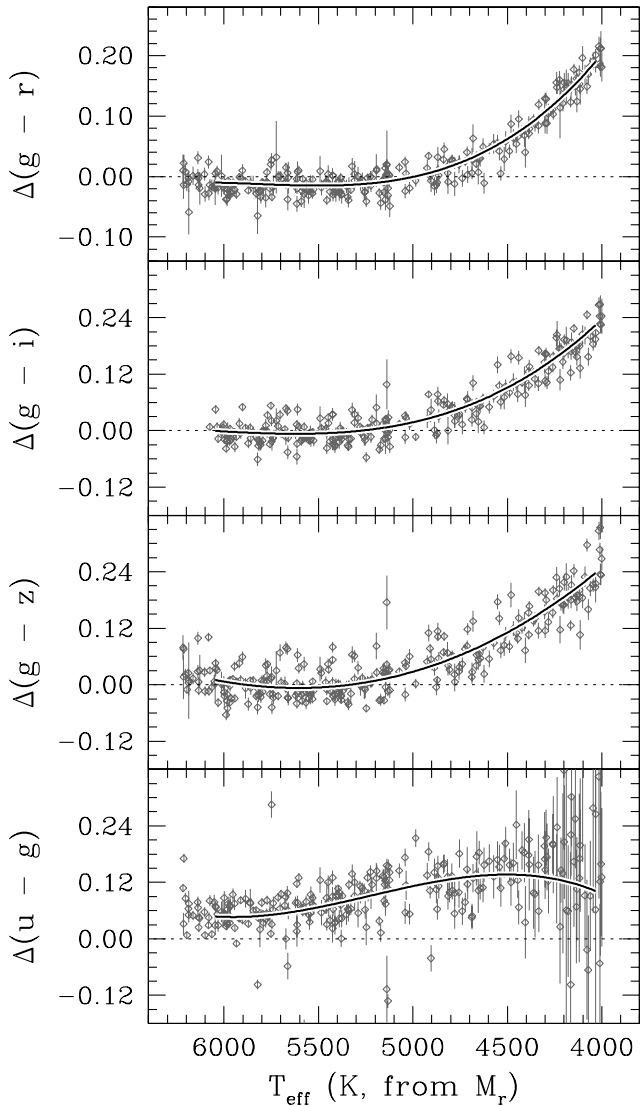}
\caption{Empirical corrections on theoretical color-$T_{\rm eff}$ relations.  Points represent
color differences of individual stars in M67 with respect to a solar-metallicity, 3.5~Gyr
old YREC+MARCS isochrone.  These stars are those selected from a photometric filtering as single
MS members based on $(g - r, r)$, $(g - i, r)$, and $(g - z, r)$.  Solid lines are empirical color
corrections to the models, which are the polynomial fits to the points in each color index.
\label{fig:empcal.m67}}
\end{figure}

Figure~\ref{fig:empcal.m67} shows color differences for individual cluster members in M67.
These stars were selected using photometric filtering of background stars and cluster
binaries on $(g - r, r)$, $(g - i, r)$, and $(g - z, r)$, independent from theoretical 
models (see Paper~I).  We used a solar-metallicity isochrone at an age of $3.5$~Gyr.
For our calibration exercise in the Johnson-Cousins-2MASS filter system \citep{pinsono:04,an:07a},
we used photometry of the Hyades and Pleiades open clusters to define empirical corrections on
the color-$T_{\rm eff}$ relations.  Similarly, we could define empirical corrections on the model
colors in $ugriz$ based on M67.

The solid lines in Figure~\ref{fig:empcal.m67} are polynomial fits to the points:
\begin{eqnarray}
\Delta (g - r) &=& 7.613 - 19.910 \theta + 17.279 \theta^2 - 4.983 \theta^3,\\
\Delta (g - i) &=& 6.541 - 16.313 \theta + 13.461 \theta^2 - 3.675 \theta^3,\\
\Delta (g - z) &=& 3.755 -  7.595 \theta +  4.535 \theta^2 - 0.672 \theta^3,\\
\Delta (u - g) &=&-7.923 + 24.131 \theta - 23.727 \theta^2 + 7.627 \theta^3,
\end{eqnarray}
where $\theta \equiv T_{\rm eff}/5040$~K.  These corrections are in the sense that the
above values should be added to the model colors, and they are strictly valid at
$4000 \leq T_{\rm eff} {\rm (K)} \leq 6000$.

It should be noted that the underlying assumptions used in these corrections are
somewhat different from our previous work.  In \citet{pinsono:03,pinsono:04}, the main
cause of the residual difference between the model and the observed MS was attributed
to the model color-$T_{\rm eff}$ relations, since the mass-luminosity-$T_{\rm eff}$
relations were found valid over $4500 \la T_{\rm eff} {\rm (K)} \la 6000$, where the
color corrections were defined.  On the other hand, the discrepancy between the lower
MS and the models at $T_{\rm eff} \la 4500$~K in Figure~\ref{fig:empcal.m67} could also
be due to our limited knowledge of the opacities and/or the equations of state in
the models at low temperatures, in addition to a problem in the model atmospheres.

\begin{figure*}
\epsscale{1.0}
\plotone{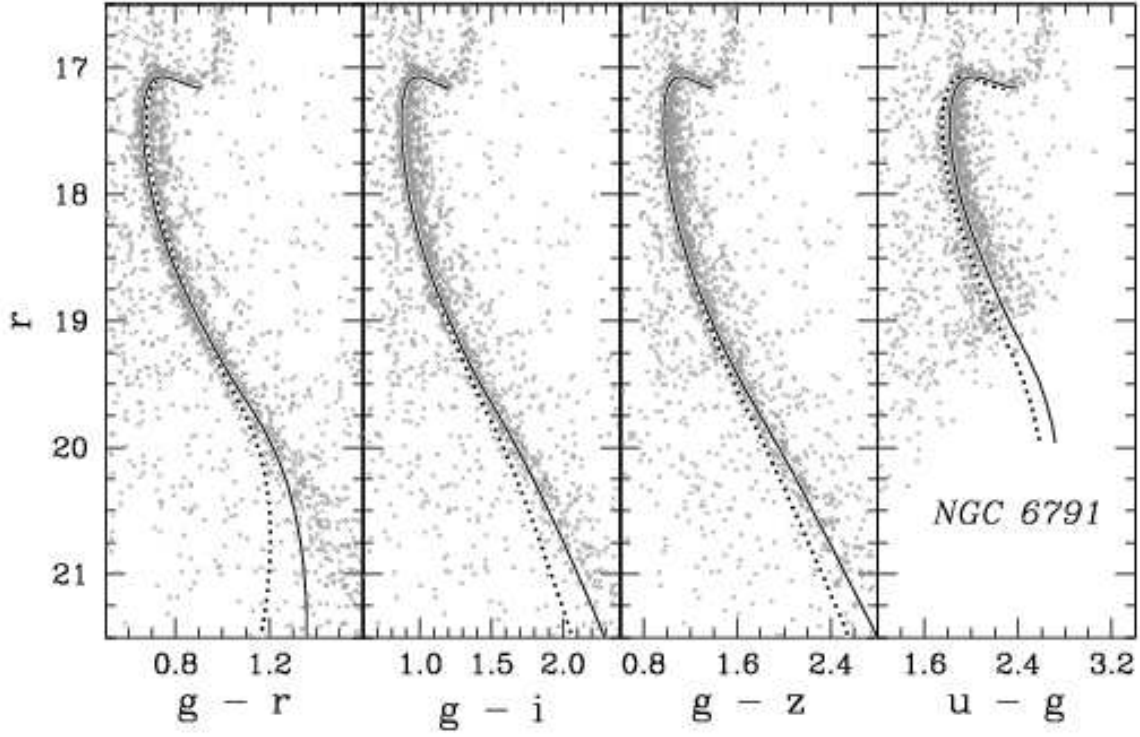}
\caption{CMDs for NGC~6791.  Dotted lines are theoretical models with [Fe/H]$ = +0.4$
at the age of 10~Gyr.  Solid lines are models with empirical color-$T_{\rm eff}$ corrections
based on M67 (see Fig.~\ref{fig:empcal.m67}).
\label{fig:n6791}}
\end{figure*}

We tested the above corrections using CMDs for NGC~6791, as shown in Figure~\ref{fig:n6791}.
Dotted lines are YREC+MARCS isochrones without corrections, and solid lines are those
with M67-based color-$T_{\rm eff}$ corrections.  Although [Fe/H]-dependent corrections
can be derived using NGC~6791, empirical corrections based on the solar metallicity
cluster result in a reasonably good agreement with the observed MS for the super-solar
metallicity cluster.  The calibrated isochrones appear to deviate from the observed MS at the
faint end, but the location of the observed MS becomes uncertain as well, due to the
increased scatter.  Because of a paucity of clusters at $-1 \la {\rm [Fe/H]} \la 0$
in the sample, the lower metallicity limit of the color corrections is not well defined.

\section{Distances and Ages of Globular Clusters}\label{sec:application}

The ages of Galactic globular clusters have been extensively discussed in the literature,
because of their importance in setting a lower limit on the age of the universe, and their
implications for the formation of the Galactic halo
\citep[e.g.,][]{sandage:70,searle:78,chaboyer:98,rosenberg:99,vandenberg:00,salaris:02,deangeli:05}.
Recent age estimates for the Galactic globular clusters are typically found between
$\sim10$~Gyr and $\sim15$~Gyr based on various techniques, including MSTO fitting
\citep[e.g.,][]{gratton:97,chaboyer:98,vandenberg:00}, termination of the white dwarf
cooling sequence \citep[e.g.][]{renzini:96,zoccali:01,hansen:02}, luminosity function
\citep[e.g.][]{jimenez:98}, and the mass-age relation for an eclipsing binary system
\citep{chaboyer:02}.  Although relative age estimates for globular clusters have been
well studied \citep[e.g.,][]{stetson:96,rosenberg:99}, estimating absolute ages is
a more difficult problem, primarily because of the uncertainty in distances to globular
clusters \citep[e.g.,][]{vandenberg:96}.

In this section, we estimate distances to our sample globular clusters using the
MS-fitting technique without employing empirical color corrections.  We then use the
derived distance to estimate cluster ages from the MSTO.  We test the validity
of deriving these quantities from the SDSS photometric database by comparing our values
with those in the literature using other methods.  In particular, we show that our
distances using YREC+MARCS models are in good agreement with the {\it Hipparcos}-based
subdwarf-fitting distance estimates.

\subsection{Distances to Globular Clusters}\label{sec:distance}

We obtained distances to globular clusters from CMDs with three color indices ($g - r$,
$g - i$, and $g - z$).  We did not use $u - g$ because of the steep MS slope in
the $(u - g, r)$ plane (Fig.~\ref{fig:all.ug}).  The utility of the multi-color
fitting technique has been demonstrated in \citet{an:07a,an:07b}.  With a wide wavelength
baseline, stellar properties are not only accurately constrained, but systematic
errors can be estimated.

\begin{deluxetable*}{lccrrrrc}
\tablewidth{0pt}
\tabletypesize{\scriptsize}
\tablecaption{MS-Fitting Distances\label{tab:msfit}}
\tablehead{
  \colhead{} &
  \colhead{GC} &
  \colhead{Reference} &
  \multicolumn{5}{c}{$(m - M)_0$} \\
  \cline{4-8}
  \colhead{Cluster} &
  \colhead{[Fe/H]} &
  \colhead{SDSS Run\tablenotemark{a}} &
  \colhead{$(g - r, r)$} &
  \colhead{$(g - i, r)$} &
  \colhead{$(g - z, r)$} &
  \colhead{Average} &
  \colhead{$\sigma_{(m - M)}$}
}
\startdata
M15      & KI03    & 2566 & $15.061\pm 0.004$ & $15.061\pm 0.007$ &  $14.930\pm 0.039$  &  $15.060\pm 0.053$ & $0.092$ \nl
         & KI03    & 1739 & $15.150\pm 0.016$ & $15.142\pm 0.011$ &  $15.011\pm 0.048$  &  $15.140\pm 0.053$ & $0.091$ \nl
         & CG97    & 2566 & $15.101\pm 0.007$ & $15.094\pm 0.008$ &  $14.958\pm 0.043$  &  $15.096\pm 0.056$ & $0.098$ \nl
         & CG97    & 1739 & $15.190\pm 0.019$ & $15.176\pm 0.014$ &  $15.038\pm 0.052$  &  $15.175\pm 0.056$ & $0.097$ \nl
M92      & KI03    & 4682 & $14.619\pm 0.019$ & $14.577\pm 0.013$ &  $14.605\pm 0.016$  &  $14.599\pm 0.012$ & $0.021$ \nl
         & KI03    & 5327 & $14.745\pm 0.013$ & $14.714\pm 0.006$ &  $14.549\pm 0.007$  &  $14.655\pm 0.062$ & $0.107$ \nl
         & CG97    & 4682 & $14.654\pm 0.017$ & $14.604\pm 0.012$ &  $14.628\pm 0.011$  &  $14.624\pm 0.015$ & $0.026$ \nl
         & CG97    & 5327 & $14.780\pm 0.014$ & $14.741\pm 0.007$ &  $14.571\pm 0.006$  &  $14.656\pm 0.070$ & $0.122$ \nl
M13      & KI03    & 3225 & $14.357\pm 0.007$ & $14.346\pm 0.003$ &  $14.326\pm 0.004$  &  $14.341\pm 0.009$ & $0.016$ \nl
         & KI03    & 3226 & $14.393\pm 0.006$ & $14.326\pm 0.005$ &  $14.316\pm 0.005$  &  $14.340\pm 0.024$ & $0.042$ \nl
         & CG97    & 3225 & $14.441\pm 0.010$ & $14.417\pm 0.006$ &  $14.392\pm 0.008$  &  $14.414\pm 0.014$ & $0.025$ \nl
         & CG97    & 3226 & $14.477\pm 0.010$ & $14.398\pm 0.006$ &  $14.382\pm 0.005$  &  $14.400\pm 0.032$ & $0.056$ \nl
M3       & KI03    & 4646 & $14.822\pm 0.011$ & $14.884\pm 0.003$ &  $14.851\pm 0.028$  &  $14.879\pm 0.026$ & $0.045$ \nl
         & KI03    & 4649 & $14.995\pm 0.006$ & $14.981\pm 0.004$ &  $14.938\pm 0.032$  &  $14.985\pm 0.020$ & $0.034$ \nl
         & CG97    & 4646 & $14.921\pm 0.009$ & $14.971\pm 0.005$ &  $14.939\pm 0.034$  &  $14.959\pm 0.018$ & $0.031$ \nl
         & CG97    & 4649 & $15.093\pm 0.010$ & $15.068\pm 0.009$ &  $15.025\pm 0.039$  &  $15.078\pm 0.023$ & $0.039$ \nl
M5       & KI03    & 1458 & $14.329\pm 0.006$ & $14.239\pm 0.002$ &  $14.294\pm 0.007$  &  $14.251\pm 0.037$ & $0.063$ \nl
         & KI03    & 2327 & $14.353\pm 0.009$ & $14.287\pm 0.005$ &  $14.247\pm 0.003$  &  $14.265\pm 0.038$ & $0.066$ \nl
         & CG97    & 1458 & $14.426\pm 0.010$ & $14.326\pm 0.002$ &  $14.376\pm 0.012$  &  $14.331\pm 0.043$ & $0.074$ \nl
         & CG97    & 2327 & $14.450\pm 0.012$ & $14.373\pm 0.007$ &  $14.328\pm 0.006$  &  $14.360\pm 0.039$ & $0.068$ \nl
M71      & KI03 & \nodata & $12.846\pm 0.022$ & $12.786\pm 0.031$ &  $12.898\pm 0.045$  &  $12.836\pm 0.033$ & $0.057$ \nl
         & CG97 & \nodata & $12.946\pm 0.028$ & $12.875\pm 0.034$ &  $12.980\pm 0.047$  &  $12.928\pm 0.031$ & $0.054$ \nl
M67      & \nodata & 5935 & $ 9.633\pm 0.014$ & $ 9.584\pm 0.014$ &  $ 9.572\pm 0.011$  &  $ 9.592\pm 0.019$ & $0.033$ \nl
         & \nodata & 5972 & $ 9.680\pm 0.021$ & $ 9.578\pm 0.013$ &  $ 9.601\pm 0.013$  &  $ 9.604\pm 0.033$ & $0.057$ \nl
         & \nodata & 6004 & $ 9.637\pm 0.114$ & $ 9.588\pm 0.014$ &  $ 9.620\pm 0.014$  &  $ 9.615\pm 0.014$ & $0.025$ \nl
NGC~6791 & \nodata & 5416 & $13.005\pm 0.021$ & $12.944\pm 0.015$ &  $12.932\pm 0.014$  &  $12.951\pm 0.024$ & $0.041$ \nl
         & \nodata & 5403 & $12.930\pm 0.016$ & $12.954\pm 0.014$ &  $13.013\pm 0.019$  &  $12.960\pm 0.025$ & $0.043$ \nl
         & \nodata & 6177 & $13.000\pm 0.019$ & $12.962\pm 0.014$ &  $12.971\pm 0.015$  &  $12.974\pm 0.012$ & $0.020$
\enddata
\tablenotetext{a}{SDSS runs selected for the local photometric zero points.}
\end{deluxetable*}

Table~\ref{tab:msfit} lists the derived distance moduli for our sample clusters,
using fiducial sequences on $(g - r, r)$, $(g - i, r)$, and $(g - z, r)$.  We restricted
our fits to $\Delta M_r \approx 1$~mag below the MSTO, to minimize any age dependency.
For the globular clusters, we list four different sets of distance estimates, on two
different metallicity scales, and with two different sets of photometric zero points
(\S~\ref{sec:comparison}).  We computed the size of the error in the average distance
modulus from the propagation of a fitting error (shape mismatch) or from a dispersion
of distance moduli from three CMDs, whichever is larger (see below for other systematic
errors).  We found, however, that the latter is always larger than the former by
approximately a factor of four.  The last column lists a standard deviation of the
distance moduli from three CMDs, indicating that distances from individual CMDs are
consistent with each other to better than $\sim5\%$.  In Table~\ref{tab:msfit}, we also
included results for open clusters (M67 and NGC6791) using purely theoretical models without
empirical corrections (\S~\ref{sec:emp}) at $r < 16.4$~mag and $r < 21.0$~mag for M67 and
NGC~6791, respectively.

\begin{deluxetable*}{lccccccccc}
\tabletypesize{\scriptsize}
\tablewidth{0pt}
\tablecaption{Systematic Errors in Distance Moduli for Individual Clusters\label{tab:sys}}
\tablehead{
  \colhead{Source of} &
  \colhead{} &
  \multicolumn{8}{c}{$\sigma_{(m - M)}$} \\
  \cline{3-10}
  \colhead{Error} &
  \colhead{$\Delta$ Quantity} &
  \colhead{M15} &
  \colhead{M92} &
  \colhead{M13} &
  \colhead{M3} &
  \colhead{M5} &
  \colhead{M71} &
  \colhead{M67} &
  \colhead{NGC~6791}
}
\startdata
Internal                   &\nodata   &$\pm0.053$&$\pm0.012$&$\pm0.009$&$\pm0.026$&$\pm0.037$&$\pm0.033$&$\pm0.019$&$\pm0.024$ \nl
Phot.\ Calibration         &$1\%-2\%$ &$\pm0.040$&$\pm0.028$&$\pm0.000$&$\pm0.053$&$\pm0.007$&$\pm0.026$&$\pm0.012$&$\pm0.012$ \nl
$R_V$                      &$\pm0.2$  &$\pm0.009$&$\pm0.002$&$\pm0.001$&$\pm0.002$&$\pm0.008$&$\pm0.040$&$\pm0.003$&$\pm0.006$ \nl
$E(B - V)$                 &$20\%$    &$\pm0.089$&$\pm0.013$&$\pm0.043$&$\pm0.008$&$\pm0.019$&$\pm0.055$&$\pm0.010$&$\pm0.032$ \nl
$\log{t}$                  &$15\%$    &$\pm0.014$&$\pm0.009$&$\pm0.007$&$\pm0.014$&$\pm0.039$&$\pm0.003$&$\pm0.007$&$\pm0.008$ \nl
$[\alpha/{\rm Fe}]$        &$\pm0.1$  &$\pm0.008$&$\pm0.005$&$\pm0.017$&$\pm0.011$&$\pm0.020$&$\pm0.015$&$\pm0.000$&$\pm0.000$ \nl
${\rm [Fe/H]}$ &$\pm0.1$\tablenotemark{a} &$\pm0.018$&$\pm0.013$&$\pm0.037$&$\pm0.040$&$\pm0.040$&$\pm0.046$&$\pm0.012$&$\pm0.027$ \nl
Total                      &\nodata   &$\pm0.114$&$\pm0.037$&$\pm0.060$&$\pm0.074$&$\pm0.073$&$\pm0.093$&$\pm0.028$&$\pm0.051$
\enddata
\tablenotetext{a}{Error in the metallicity scale between CG97 and KI03 for globular clusters.
The $\sigma_{\rm [Fe/H]} = \pm0.01$ and $\sigma_{\rm [Fe/H]} = \pm0.03$ were assumed for M67 and NGC~6791, respectively.}
\end{deluxetable*}

Table~\ref{tab:sys} lists systematic errors in distance estimation.  We used the same list
and size of systematic error sources as in Table~\ref{tab:sys2}.  An ``internal'' error was
taken from the error in the average distance modulus in Table~\ref{tab:msfit}.  Although
the adopted size of the photometric calibration errors varies from one cluster to the other,
the small size of the errors (e.g., M13) may reflect an accidental agreement of photometry
from two SDSS runs.  The error from [Fe/H] increases at higher metallicities, because the
metallicity sensitivity of colors and magnitudes increases steeply with [Fe/H].
A quadrature sum of all of these systematic errors is generally larger than the internal
systematic error.  We note that no particular source of systematic errors dominate the
total error budget in our distance determination.  Average distance moduli for individual
clusters and their total errors are listed in the last two columns of Table~\ref{tab:dist.gc},
on the CG97 and KI03 metallicity scales, respectively.

\begin{figure}
\epsscale{1.2}
\plotone{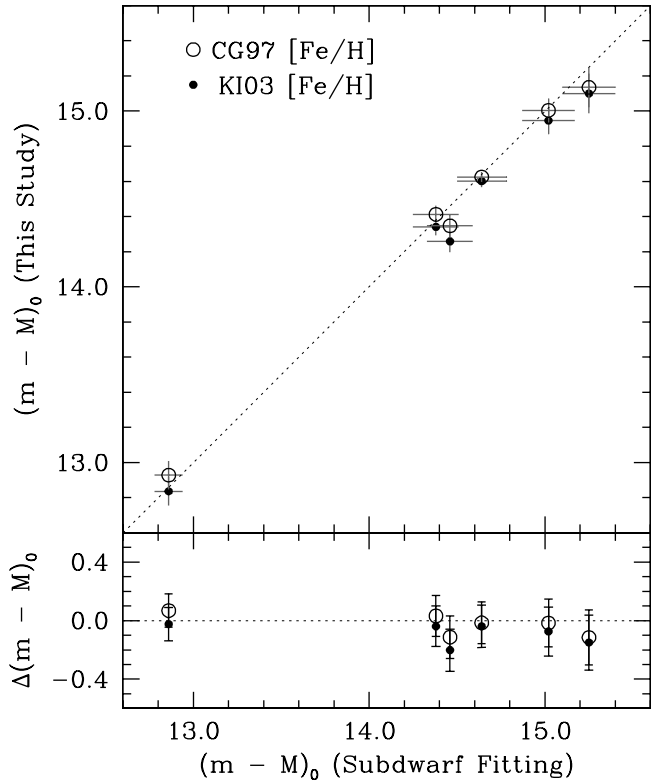}
\caption{Comparison of distance moduli for globular clusters from the {\it Hipparcos}-based
subdwarf fitting (see \S~\ref{sec:data}) and this study.  Open circles are those estimated from
the KI03 [Fe/H] values, and the closed circles are those from the CG97 [Fe/H] values.
\label{fig:hipp}}
\end{figure}

\begin{figure}
\epsscale{1.2}
\plotone{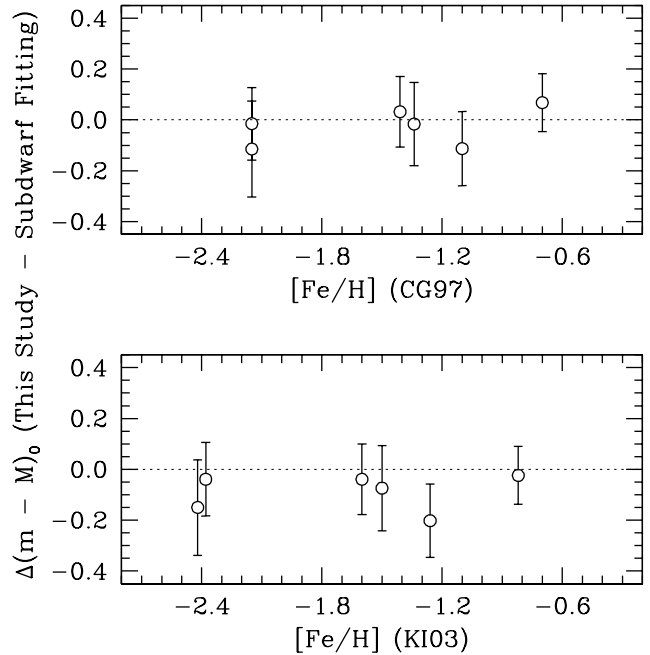}
\caption{Same as in Fig.~\ref{fig:hipp}, but as a function of metallicity.
{\it Top:} Comparison of distance moduli when the CG97 [Fe/H] values are adopted in the MS fitting.
{\it Bottom:} Comparison when the KI03 [Fe/H] values are adopted.
\label{fig:hipp.feh}}
\end{figure}

Figure~\ref{fig:hipp} compares our distance moduli for globular clusters with those based on
{\it Hipparcos} subdwarf fitting.  Open circles are those estimated using the CG97 [Fe/H]
values, while closed circles are those estimated using the KI03 [Fe/H] values.
Figure~\ref{fig:hipp.feh} shows these differences, as a function of metallicity, on the two
different [Fe/H] scales.  Error bars are the quadrature sum of the errors from both our values
and those from the subdwarf-fitting studies.  The average difference between these two studies
is $\Delta (m - M)_0 = -0.02\pm0.07$ on the CG97 [Fe/H] scale, and
$\Delta (m - M)_0 = -0.08\pm0.07$ on the KI03 scale.  In both cases, our distances are
consistent with subdwarf-fitting distances.  Furthermore, the agreement between subdwarf-fitting
and this study indicates that our error estimate ($\sigma_{(m-M)} \sim 0.03$--$0.11$~mag) for
individual clusters is reasonable.  In fact, the agreement between the two distance
determinations restates our earlier conclusion that the YREC+MARCS models are in good agreement
with cluster fiducial sequences.

The distance modulus of M71 from
\citet{grundahl:02} is $0.33\pm0.17$~mag smaller than the value in \citet{reid:98}, although
these studies assumed the same $E(B - V) \approx 0.28$.  Our best-fitting distance modulus of
this cluster, $(m - M)_0 = 12.86\pm0.08$ (KI03) or $12.96\pm0.08$ (CG97), are in better
agreement with the \citeauthor{grundahl:02} value.  Our cluster distance modulus is also
in agreement with $(m - M)_0 = 12.76\pm0.18$ from \citet{geffert:00}.
In addition, the average value from the Harris compilation becomes $(m - M)_0 = 12.92$
when we use $E(B - V) = 0.28$, rather than his adopted $E(B - V) \approx 0.32$.

C00 paid particular attention to M92, not only because of its importance as one of the
most metal-poor Galactic clusters, but also because of the large discrepancies in
its reported distance in the literature.  Two factors make measuring the cluster's
distance difficult.  First, there are only a few metal-poor subdwarfs with
${\rm [Fe/H]} \la -2$ in the {\it Hipparcos} catalog.  Secondly, the two independent
photometric studies \citep{heasley:86,stetson:88} that provided $BV$ photometry have
$0.03$~mag differences in colors.
As shown in Table~\ref{tab:dist.gc}, with independent photometric data and a wide wavelength
coverage in $griz$, we found that our distance to M92 is in good agreement with the C00
distance, independent of the metallicity scale used.

Here we neglected the effects of unresolved binaries.  In \citet{an:07b} we have performed
extensive simulations of unresolved binaries in clusters and inspected their influence on
the estimation of a MS-fitting distance.  After applying photometric filtering, as was
done in the present paper for our cluster sample, we found that unresolved binaries can make
a MS look brighter by $\sim0.007$~mag for a 40\% binary fraction\footnote{Binary fraction is
defined as the number of binaries divided by the total number of systems.}.  The observed
binary fraction in globular clusters is typically less than $20\%$ \citep[e.g.,][and references
therein]{sollima:07,davis:08}, which makes the influence of unresolved binaries even smaller.

\subsection{Cluster Ages}

\begin{figure*}
\epsscale{1.0}
\plotone{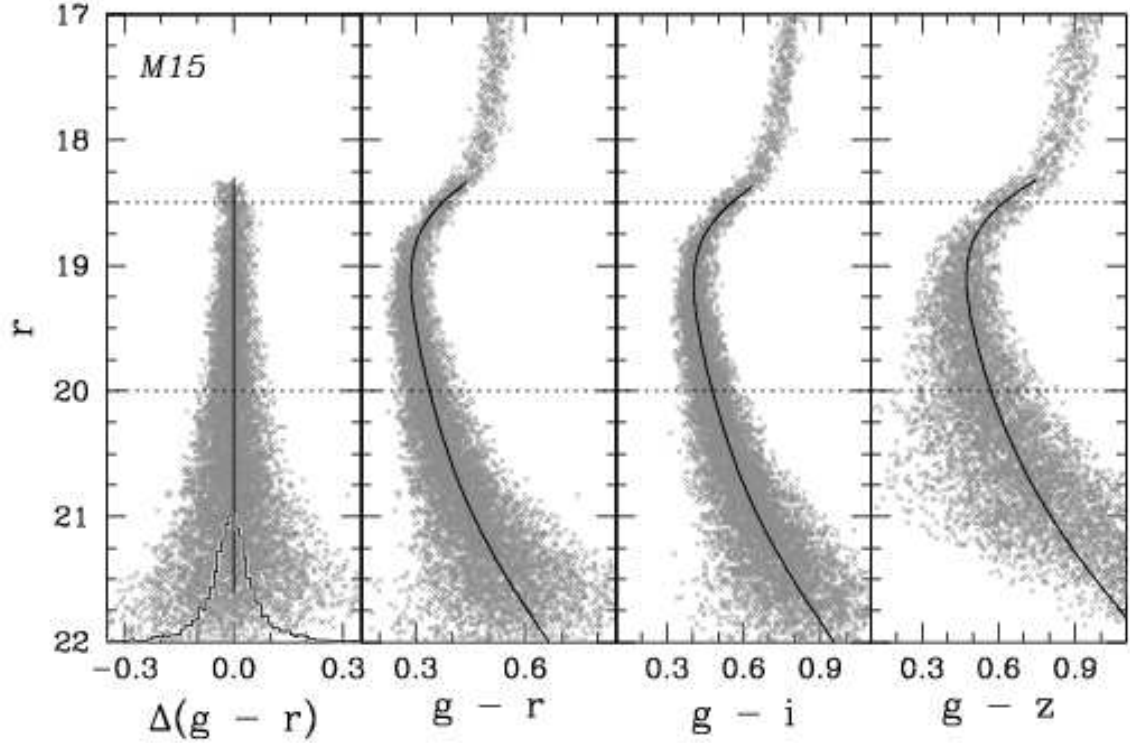}
\caption{CMDs for M15 with YREC+MARCS models at the best-fitting distance and age
({\it solid line}).  Models without empirical color corrections are used.
Dotted, horizontal lines represent a magnitude range where the
isochrone fit was performed to estimate the cluster age.  Points are stars that were
remained after the photometric filtering (see text).  The leftmost panel shows the
fitting residuals in $(g - r, r)$.  Histograms are the number distribution of the residuals,
with an arbitrary scale on the vertical axis.
\label{fig:obs.m15}}
\end{figure*}

\begin{figure*}
\epsscale{1.0}
\plotone{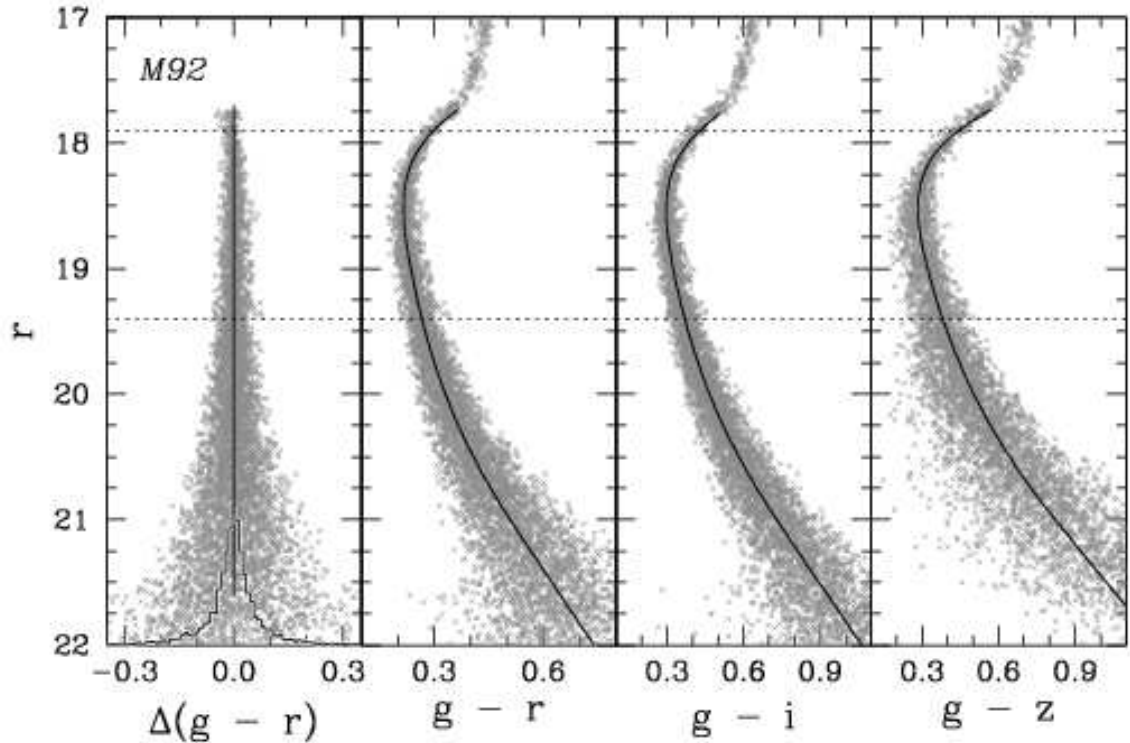}
\caption{Same as in Fig.~\ref{fig:obs.m15}, but for M92.
\label{fig:obs.m92}}
\end{figure*}

\begin{figure*}
\epsscale{1.0}
\plotone{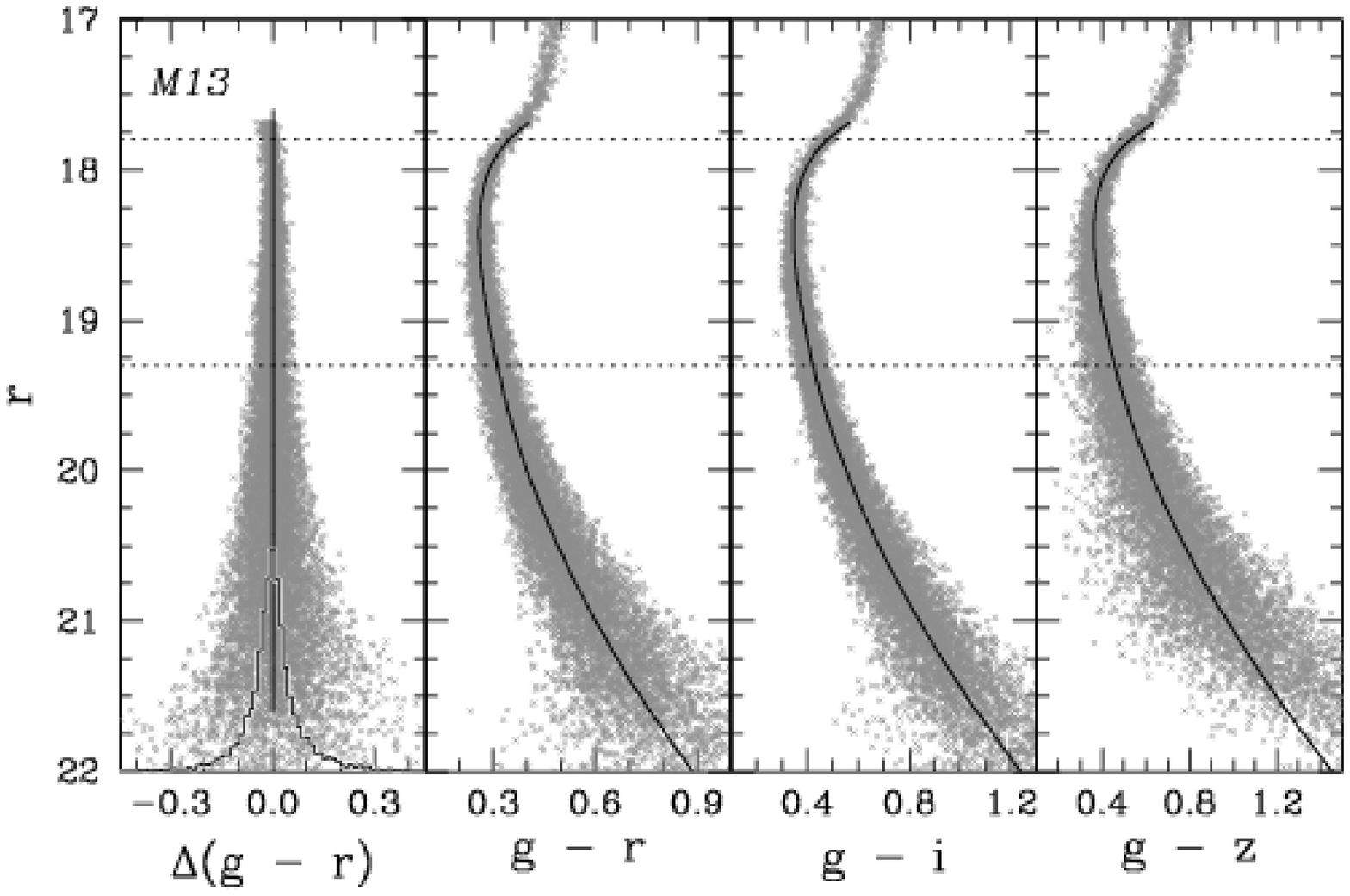}
\caption{Same as in Fig.~\ref{fig:obs.m15}, but for M13.
\label{fig:obs.m13}}
\end{figure*}

\begin{figure*}
\epsscale{1.0}
\plotone{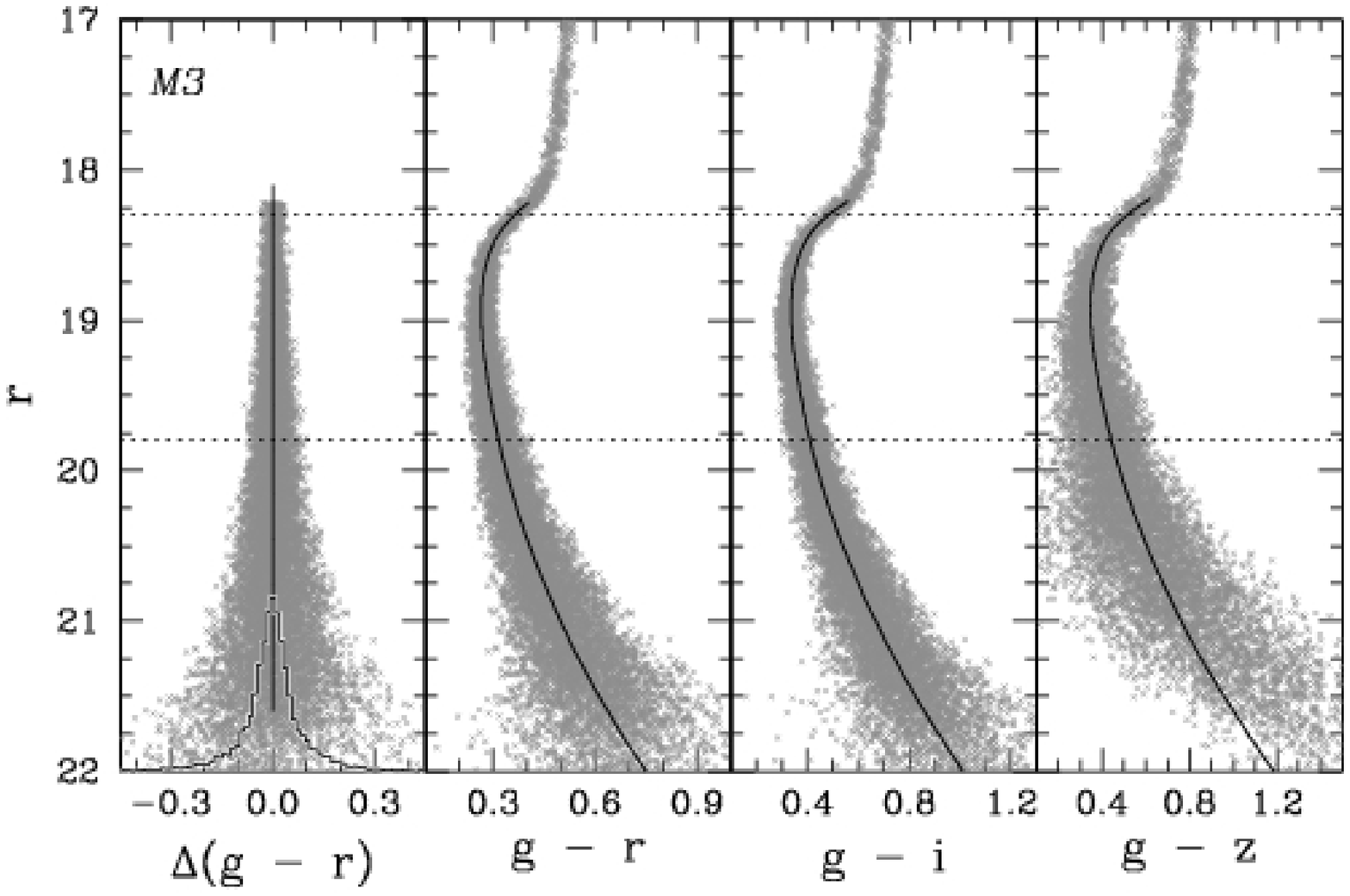}
\caption{Same as in Fig.~\ref{fig:obs.m15}, but for M3.
\label{fig:obs.m3}}
\end{figure*}

\begin{figure*}
\epsscale{1.0}
\plotone{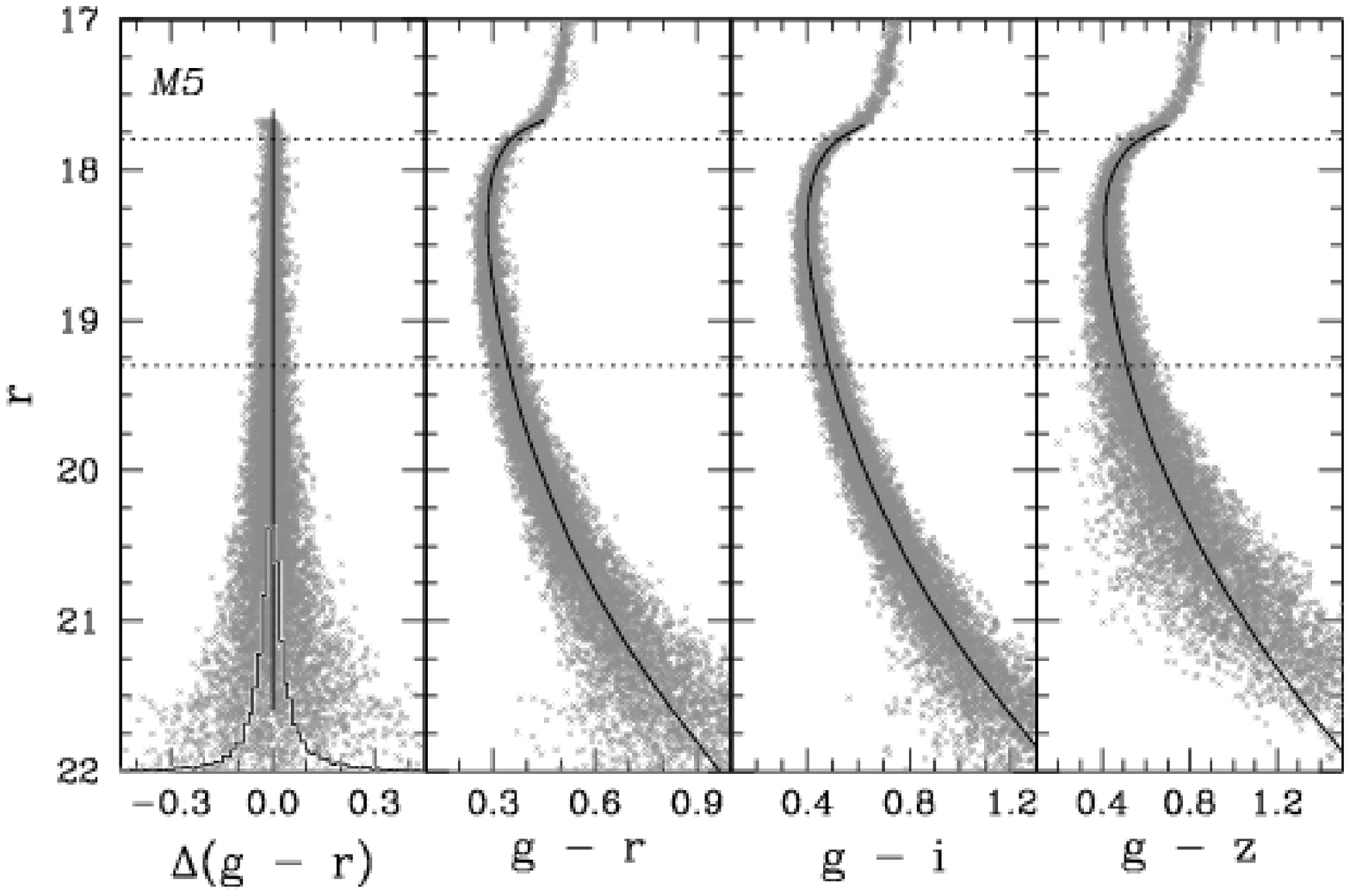}
\caption{Same as in Fig.~\ref{fig:obs.m15}, but for M5.
\label{fig:obs.m5}}
\end{figure*}

We have derived cluster ages for our sample using MSTO stars, based on the MS-fitting
distances to clusters described above.  Figures~\ref{fig:obs.m15}--\ref{fig:obs.m5}
display $(g - r, r)$, $(g - i, r)$, and $(g - z, r)$ CMDs, respectively, for five globular
clusters in our sample, with best-fitting stellar isochrones (solid lines, see below).
We fit directly to the CMDs, instead of using the cluster fiducial sequences, because
fiducial sequences can be inaccurate near the MSTO (where curvatures are largest on CMDs).

The first panel in Figures~\ref{fig:obs.m15}--\ref{fig:obs.m5} exhibits residuals
on $(g - r, r)$ with respect to a best-fitting model.  The points in the remaining three
panels are those after applying cuts based on the $\chi$, sharp, and separation indices,
as well as the photometric filtering (see Paper~I for more details).  The photometric filtering
routine statistically eliminates cluster non-members and unresolved binaries, independent
of the stellar isochrones.  However, we note that cuts based on the separation index already
make a cluster fiducial sequence sufficiently narrow for well-populated globular clusters such
as those in our sample.

To find a best-fitting age on each CMD, we minimized
\begin{equation}
\chi^2_{\rm tot} = \sum_{i=1}^N \frac{(X_i - X_m)^2}{\sigma_{X,i}^2 + (\gamma \sigma_{r,i})^2},
\end{equation}
at the best-fitting distance derived simultaneously using MS stars.  Here, $X_i$ and
$\sigma_{X,i}$ are a color and its error for the $i^{th}$ point, respectively.  The
$\sigma_{r,i}$ is an error in $r$-band magnitude, weighted by the inverse slope
of the isochrone $\gamma$.  The $X_m$ is a model color at the star's $r$-band magnitude.
We used stars that are within $\sim0.5$~mag in $r$ on the bright side of the bluest
point of the MSTO and $\sim1.0$~mag on the faint side.  The bright magnitude limit was
set by the fact that our evolutionary model computations do not reach as far beyond the
hydrogen exhaustion in the core.  On the red side of subgiant branch (SGB), assumptions
on the mixing length and/or the model color-$T_{\rm eff}$ relations become uncertain as
well.  The faint end was set at this level, since the MS is insensitive to the
age below this limit.  Dotted horizontal lines in Figures~\ref{fig:obs.m15}--\ref{fig:obs.m5}
show these fitting ranges.

Solid lines in Figures~\ref{fig:obs.m15}--\ref{fig:obs.m5} represent our
best-fitting models for individual CMDs, using KI03 [Fe/H] values.  The
$g - r$ residuals in the first panels show that fits are excellent from
the lower MS to the MSTO.  Histograms are also shown to visualize
a peaked distribution of color residuals from the model.  Reduced $\chi^2$
values were between $\sim0.9$ and $\sim1.2$ for approximately 2000 stars
on each CMD.

\begin{deluxetable}{lcc}
\tablewidth{0pt}
\tablecaption{MSTO Ages of Globular Clusters\label{tab:age}}
\tablehead{
  \colhead{} &
  \multicolumn{2}{c}{Age (Gyr)} \\
  \cline{2-3}
  \colhead{Cluster} &
  \colhead{${\rm [Fe/H]}_{\rm KI03}$} &
  \colhead{${\rm [Fe/H]}_{\rm CG97}$}
}
\startdata
M15 & $13.9\pm2.5$ & $12.8\pm2.3$ \nl
M92 & $14.4\pm0.9$ & $13.5\pm0.9$ \nl
M13 & $14.3\pm1.1$ & $12.3\pm1.0$ \nl
M3  & $13.3\pm1.4$ & $11.4\pm1.1$ \nl
M5  & $12.2\pm1.3$ & $10.5\pm1.2$
\enddata
\end{deluxetable}

\begin{deluxetable*}{lcccccc}
\tablewidth{0pt}
\tablecaption{Systematic Errors in the Ages of Globular Clusters\label{tab:sys.age}}
\tablehead{
  \colhead{Source of} &
  \colhead{} &
  \multicolumn{5}{c}{$\sigma_{\log{t}}$} \\
  \cline{3-7}
  \colhead{Error} &
  \colhead{$\Delta$ Quantity} &
  \colhead{M15} &
  \colhead{M92} &
  \colhead{M13} &
  \colhead{M3} &
  \colhead{M5}
}
\startdata
Internal                &\nodata   &$\pm0.024$ &$\pm0.014$ &$\pm0.009$ &$\pm0.012$ &$\pm0.027$ \nl
Phot.\ Calibration      &$1\%-2\%$ &$\pm0.022$ &$\pm0.015$ &$\pm0.006$ &$\pm0.024$ &$\pm0.004$ \nl
$(m - M)_0$             & Internal &$\pm0.003$ &$\pm0.003$ &$\pm0.001$ &$\pm0.003$ &$\pm0.002$ \nl
$R_V$                   &$\pm0.2$  &$\pm0.009$ &$\pm0.003$ &$\pm0.002$ &$\pm0.000$ &$\pm0.002$ \nl
$E(B - V)$              &$20\%$    &$\pm0.061$ &$\pm0.012$ &$\pm0.019$ &$\pm0.007$ &$\pm0.019$ \nl
$[\alpha/{\rm Fe}]$     &$\pm0.1$  &$\pm0.010$ &$\pm0.011$ &$\pm0.023$ &$\pm0.029$ &$\pm0.028$ \nl
${\rm [Fe/H]}$          &$\pm0.1$  &$\pm0.018$ &$\pm0.015$ &$\pm0.032$ &$\pm0.033$ &$\pm0.034$ \nl
Total                   &\nodata   &$\pm0.073$ &$\pm0.030$ &$\pm0.045$ &$\pm0.052$ &$\pm0.055$ \nl
                        &          &$(\pm16.8\%)$&$(\pm7.0\%)$&$(\pm10.4\%)$&$(\pm12.0\%)$&$(\pm12.7\%)$
\enddata
\end{deluxetable*}

Our best-fitting ages are tabulated in Table~\ref{tab:age}.  The errors are the
quadrature sums of various systematic errors in the estimation, as listed in
Table~\ref{tab:sys.age}.  The same sizes of systematic errors were adopted as in
Tables~\ref{tab:sys2} and \ref{tab:sys}.  The first row in Table~\ref{tab:sys.age}
shows the scatter in the age estimation from three CMDs ($g - r$, $g - i$, and
$g - z$), which describes an internal systematic error of $\sim0.5$~Gyr.  Since
the distance and age estimates are correlated with each other, we evaluated
the effects of various systematic errors, while simultaneously solving
for both quantities.  In Table~\ref{tab:sys.age} we separately tabulated the
contribution from an internal error component in distance (see Table~\ref{tab:sys}).
We took the difference in age estimates from the CG97 and KI03 [Fe/H] scales as
a $2\sigma$ error.  Our error estimates indicate that we can obtain cluster ages
with an accuracy of $\sim10\%$--$15\%$ using multiple color indices in SDSS.

\begin{figure}
\epsscale{1.2}
\plotone{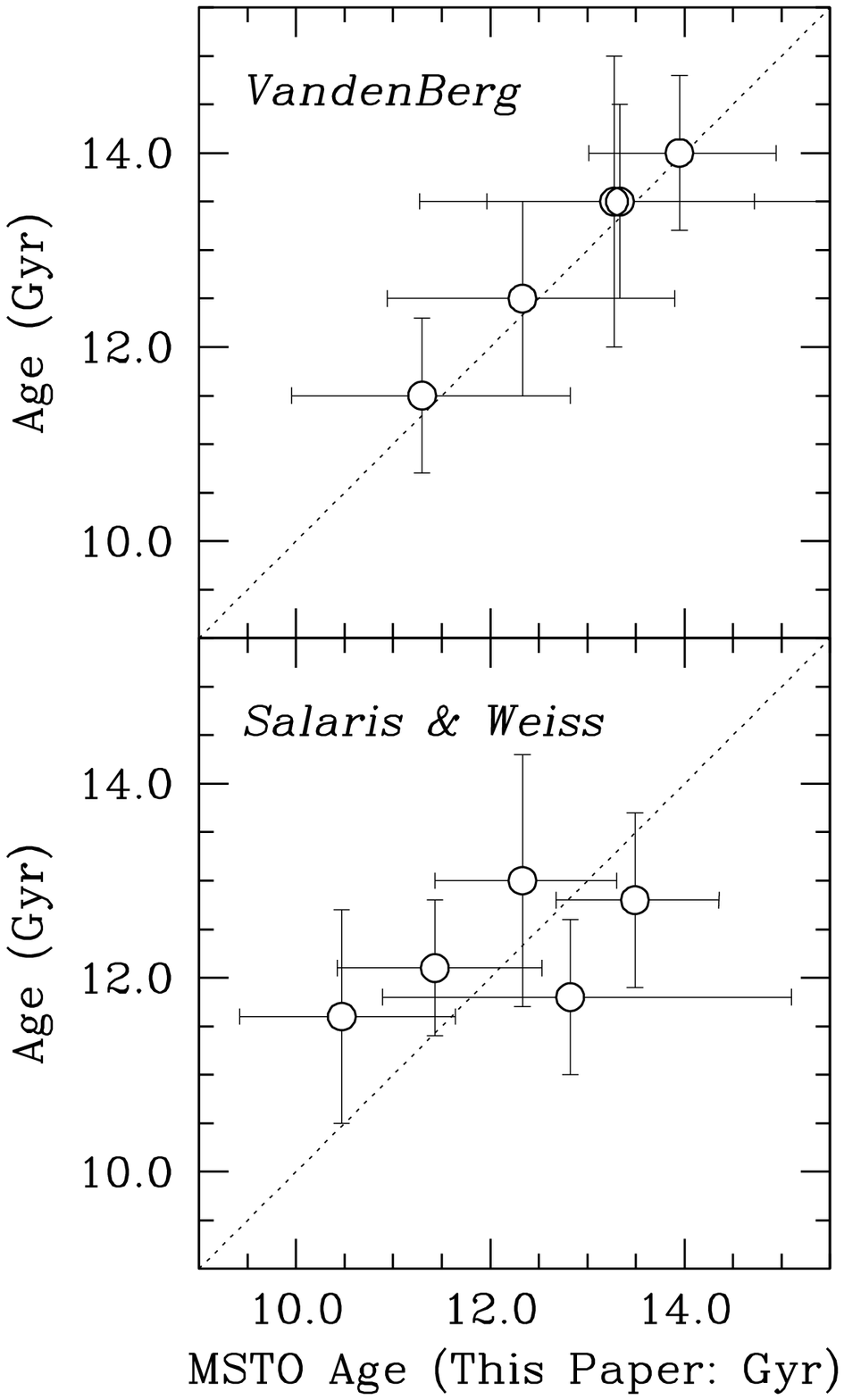}
\caption{{\it Top:} Comparison of globular cluster age estimates in \citet{vandenberg:00}
and those in this study.  The latter are the average values computed from the CG97 and KI03
[Fe/H] values, while the former age estimates are based on those [Fe/H] values adopted in
that study (see text).  {\it Bottom:} Comparison between \citet{salaris:02} and this study.
Both of these age estimates are based on the CG97 [Fe/H] values.
\label{fig:logt3}}
\end{figure}

The accuracy of our age estimates can be further checked through a comparison
with previous studies.  Figure~\ref{fig:logt3} shows comparisons of our age
estimates with those in two recent studies: \citet[][{\it top}]{vandenberg:00}
and \citet[][{\it bottom}]{salaris:02}.  \citet{vandenberg:00} estimated cluster
ages from the luminosity difference between his theoretical HB models and two
points on the cluster's MS and SGB, which are $0.05$~mag redder in $B - V$
than the bluest point of the MSTO.  Because the cluster [Fe/H] values adopted
in \citet{vandenberg:00} were generally found to lie between the CG97 or
KI03 values,\footnote{\citet{vandenberg:00} adopted [Fe/H]$ = -2.30$ for M15
and M92, [Fe/H]$ = -1.60$ for M3 and M13, and [Fe/H]$ = -1.40$ for M5.} we used
the average of our age estimates at the CG97 and KI03 [Fe/H] values in the comparison.
On the other hand, \citet{salaris:02} derived ages for M3 and M15 from the $V$
magnitude difference between Zero-Age HB and MSTO, but employed relative age
estimates for other clusters in our sample from the color difference in $B - V$
or $V - I_C$ between the MSTO and the base of RGB.  They presented ages from both
the \citet{zinn:84} and CG97 [Fe/H] scales, but only those from the latter [Fe/H]
scale are shown in Figure~\ref{fig:logt3}.

There is excellent agreement between our age estimates and those in
\citet{vandenberg:00} (Fig.~\ref{fig:logt3}), even though they employed
independent photometry, photometric filters, stellar isochrones, and different
methods of estimating cluster distances and ages.  In addition to the above
two studies, C00 derived ages for 9 globular clusters by comparing MSTO
luminosities with model isochrones \citep{straniero:97}.  They obtained
14.8, 12.6, and 11.2~Gyrs for M92, M13, and M5, respectively, which are in
agreement with our values.
The outcome of the comparisons between ours and the preceding studies
demonstrates that cluster ages derived with the SDSS $ugriz$ system are
reliable, and that our error estimates on age are reasonable. 

However, we also noted above that theoretical errors
\citep[e.g.][]{vandenberg:96,chaboyer:98,krauss:03} were not included in our
total error budget.  For example, the inclusion of the microscopic diffusion
\citep{korn:06,korn:07} can systematically reduce stellar ages by $\sim10\%$
\citep[e.g.,][]{chaboyer:01,vandenberg:02,michaud:04}.  Nevertheless, our
fitting results (Figs.~\ref{fig:obs.m15}--\ref{fig:obs.m5}) are encouraging,
showing excellent fits in all three of the color indices.  This suggests
that our mixing length calibration based on the Sun is valid for stars
in globular clusters, and that theoretical color-$T_{\rm eff}$ relations are
also good for MSTO stars, as we found for MS stars in \S~\ref{sec:comparison}.

Cluster ages have traditionally been estimated from the absolute $V$ magnitude
of the bluest point on MSTO (e.g., $\Delta V_{\rm TO}^{\rm HB}$ method).  This
was because the luminosity of the MSTO was considered to be the most robustly
predicted quantity in stellar evolutionary models \citep[e.g.,][]{bolte:95,chaboyer:96b}.
Our approach of fitting isochrones to the whole MSTO has been cautioned against
in the literature \citep[e.g.,][]{renzini:91}, primarily because of limits
on knowledge of the mixing length parameter and the color-$T_{\rm eff}$
relations used in the models.  However, the $\Delta V_{\rm TO}^{\rm HB}$ method
has its own limitations, in particular the observational difficulty of pinning
down the bluest MSTO point.  Because the MS is almost vertical in that region, the
magnitude of the MSTO cannot be better determined than $\Delta V \sim 0.1$~mag,
which then produces an error of $\sim1.5$~Gyr in age.  Subsequent studies modified the
$\Delta V_{\rm TO}^{\rm HB}$ technique in such a way that slightly redder points
than the MSTO are used as an age indicator \citep[e.g.,][]{chaboyer:96a,vandenberg:00}.
However, this method implicitly assumes that the shape of MSTO in the models
is correct, as we have in this paper.

\begin{figure}
\epsscale{1.2}
\plotone{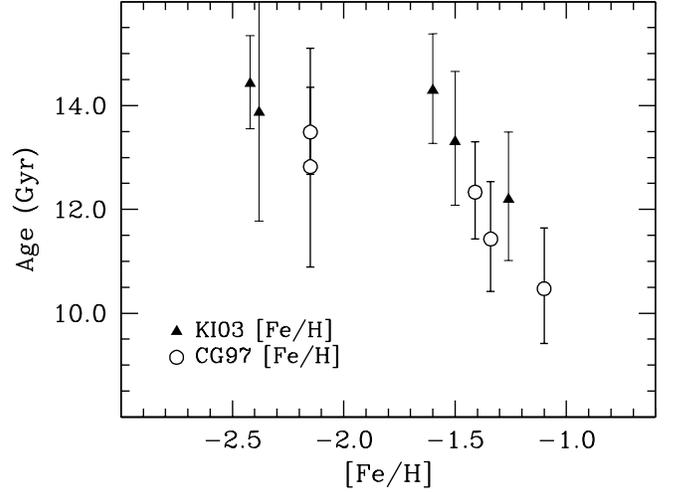}
\caption{Turn-off ages of globular clusters in this study as a function of metallicity.
Filled triangles and open circles represent the cases of adopting the KI03 or CG97 [Fe/H]
values, respectively.
\label{fig:logt}}
\end{figure}

Figure~\ref{fig:logt} shows MSTO ages for our sample globular clusters
as a function of metallicity.  The open circles are those estimated from the
CG97 [Fe/H] values, while the filled triangles are those from the KI03 values.
Error bars represent the total error in age, as listed in Table~\ref{tab:age}.
With constraints on the distance, we can inspect how stellar ages can be
affected by secondary factors, such as the cluster [Fe/H] scale.  When the
KI03 [Fe/H] values are adopted, the average age of our sample clusters is
$13.8\pm0.5$~Gyr.  Some of the clusters are even older than the inferred age
of the universe based on {\it the Wilkinson Microwave Anisotropy Probe (WMAP)}
data \citep[$\approx13.7$~Gyr,][]{spergel:03,dunkley:08}, although they are
still consistent with each other within the errors.  However, the average MSTO
age becomes $12.2\pm0.5$~Gyr when the CG97 values are used, reducing tension
between stellar and {\it WMAP}-based ages.  In addition, the age estimates
based on the CG97 scale exhibit a steeper age-metallicity relation.
Although only five clusters are included in the analysis, Figure~\ref{fig:logt}
demonstrates that a different cluster [Fe/H] scale could lead to a different
chemical enrichment scenario for the Galactic globular cluster system.

\begin{figure}
\epsscale{1.2}
\plotone{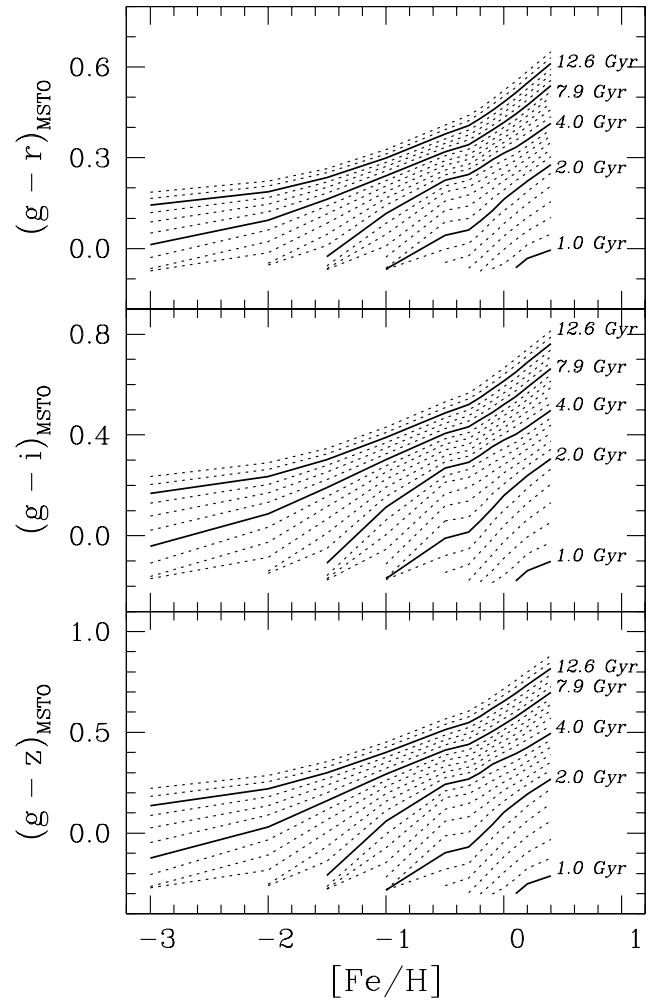}
\caption{Theoretical MSTO colors in three color indices.
\label{fig:msto}}
\end{figure}

Finally, theoretical colors of the MSTO are presented in Figure~\ref{fig:msto}
for three color indices, to aid the interpretation of CMDs in SDSS, as well as for
future imaging surveys such as the Panoramic Survey Telescope \& Rapid Response
System \citep[Pan-STARRS;][]{kaiser:02} and the Large Synoptic Survey Telescope
\citep[LSST;][]{ivezic:08b}, which will use similar $ugriz$ filter system.  No
empirical corrections were applied to the theoretical colors.

\section{Summary and Discussion}

We generated a set of stellar isochrones in the SDSS $ugriz$ system and performed
an extensive test of these models for MS stars.
We found that the models match observed cluster sequences within the errors of
the adopted cluster parameters, for CMDs using $g - r$, $g - i$, and $g - z$ as color indices.
These models correctly predict stellar
colors and magnitudes over a wide range of metallicity, given the size of errors
in distance, reddening, metallicity, and photometric calibration.  However, we
found $\sim0.05$--$0.10$~mag offsets in $u - g$, indicating a problem in the
model atmospheres for the $u$ bandpass.  More accurate tests will be possible when
future astrometric missions such as {\it Gaia} \citep{perryman:01} provide
improved distance measurements for these clusters.

On the other hand, we found a strong departure of the models from the open
cluster observations of the lower MS stars.  The problem with cool stars
has long been noted in other photometric systems, such as in Johnson-Cousins
bandpasses, and is presumably due to our limited knowledge on the opacities
and/or the equations of state used in the models.  From the observation of M67,
we proposed empirical corrections on the model color-$T_{\rm eff}$ relations to
alleviate a large distance error to the cluster from MS fitting.  Our
tests indicate that isochrones with these corrections provide an improved fit to
MS of NGC~6791.  Since colors and magnitudes of stars change rapidly with metallicity,
accurate photometry of open clusters in $ugriz$ will be useful to explore
metallicity-dependent corrections on color-$T_{\rm eff}$ relations in the future.

We also tested models from Padova+ATLAS9 \citep{girardi:04} and DSEP+PHOENIX
\citep{dotter:08} using cluster fiducial sequences as for comparison with the
YREC+MARCS models -- the models provide reasonable fits to the observations generally,
but exhibit strong departures for the lower MS at solar and super-solar metallicities.
However, Padova+ATLAS9 models do not include $\alpha$-element enhancement, prohibiting
more accurate comparisons with cluster observation.  In addition to $u - g$, we also
found a large offset in $g - z$ for DSEP+PHOENIX models.
In general, our YREC+MARCS models better match observed MSs than these models.

We derived distances and MSTO ages for our sample of Galactic globular clusters
using our theoretical stellar isochrones.  We performed isochrone fits for
three color indices ($g - r$, $g - i$, and $g - z$) over a wide wavelength
range to constrain these quantities and to assess the size of internal
systematic errors.
Our distance estimates are consistent with those from {\it Hipparcos}-based
subdwarf fitting, with an error of $\sigma_{(m-M)} \sim 0.03$--$0.11$~mag for
individual clusters.  We also found good agreement of our age estimates with
those in the literature, with $\sigma_{\rm age} \sim 10\%$--$15\%$.
Our results demonstrate the ability to derive distances and ages for stars
observed in SDSS, as well as in the future from imaging surveys such as
Pan-STARRS \citep{kaiser:02} and with LSST \citep{ivezic:08b}, which will
employ similar $ugriz$ filter systems.

\acknowledgements

We thank the referee for careful and detailed comments.
D.A.\ thanks Grant Newsham for kindly proving his isochrone interpolation code.
We thank Aaron Dotter and Brian Chaboyer for helpful discussions on the model
comparison. D.A.\ gratefully acknowledges the support from the Presidential
Fellowship in the Ohio State University.  F.D.\ is supported by ANR (ANR-06-CIS6-009-01).
T.C.B.\ is grateful for support in part by from grant AST 07-07776, and grants
PHY 02-16783 and PHY 08-22648: Physics Frontiers Center / Joint Institute for
Nuclear Astrophysics (JINA), awarded by the U.S. National Science Foundation.
Any opinion, finding, and conclusions or recommendations expressed in
this material are those of the author and do not necessarily reflect
the views of the National Science Foundation.

Funding for the SDSS and SDSS-II has been provided by the Alfred P.\ Sloan
Foundation, the Participating Institutions, the National Science Foundation,
the U.S.\ Department of Energy, the National Aeronautics and Space Administration,
the Japanese Monbukagakusho, the Max Planck Society, and the Higher Education
Funding Council for England. The SDSS Web Site is http://www.sdss.org/.

The SDSS is managed by the Astrophysical Research Consortium for the Participating
Institutions. The Participating Institutions are the American Museum of Natural
History, Astrophysical Institute Potsdam, University of Basel, University of
Cambridge, Case Western Reserve University, University of Chicago, Drexel
University, Fermilab, the Institute for Advanced Study, the Japan Participation
Group, Johns Hopkins University, the Joint Institute for Nuclear Astrophysics,
the Kavli Institute for Particle Astrophysics and Cosmology, the Korean Scientist
Group, the Chinese Academy of Sciences (LAMOST), Los Alamos National Laboratory,
the Max-Planck-Institute for Astronomy (MPIA), the Max-Planck-Institute for
Astrophysics (MPA), New Mexico State University, Ohio State University, University
of Pittsburgh, University of Portsmouth, Princeton University, the United States
Naval Observatory, and the University of Washington.

\end{document}